\documentclass{aa}
\usepackage{natbib}
\usepackage{graphicx}
\usepackage{txfonts}
\usepackage{color}
\begin{document}
   \title{Magnetic flux generation and transport 
in cool stars}

   \author{E. I\c{s}\i k$^{1,2}$
          \and
	  D. Schmitt$^{1}$
	  \and 
	  M. Sch\"ussler$^{1}$
}

   \institute{  Max-Planck-Institut f\"ur Sonnensystemforschung,
        37191 Katlenburg-Lindau, Germany 
	\and
	Department of Physics, Faculty of Science \& Letters, 
	\.Istanbul K\"ult\"ur University, Atak\"oy Campus, 
	Bak\i rk\"oy 34156, \.Istanbul, Turkey \\
             \email{e.isik@iku.edu.tr; schmitt,schuessler@mps.mpg.de}
             }
   \date{\today}

\abstract
{The Sun and other cool stars harbouring outer convection zones manifest 
 magnetic activity in their atmospheres. 
The connection between this activity and the properties of a 
deep-seated dynamo generating the magnetic flux is not well understood.}
{By employing physical models, we study
 the spatial and temporal characteristics 
 of the observable surface field for various stellar parameters.}
{We combine models for magnetic flux generation, buoyancy
instability, and transport, which encompass the entire convection zone. The 
model components are: ($i$) a thin-layer $\alpha\Omega$ dynamo at the
base of the convection zone; 
($ii$) buoyancy instabilities and the rise of flux tubes through the
convection zone in 3D, 
which provides 
a physically consistent determination of emergence latitudes and tilt angles; 
and ($iii$) horizontal flux transport at the surface. }
{For solar-type stars and rotation periods longer than about 10 days, 
the latitudinal dynamo waves generated by the deep-seated $\alpha\Omega$ 
dynamo are faithfully reflected 
by the surface distribution of magnetic flux. For rotation periods
of the order of two days, however, Coriolis acceleration of rising flux 
loops leads 
to surface flux emergence at much higher latitudes than the dynamo
waves at the bottom of the 
convection zone reach. A similar result is found for a K0V star with a rotation
period of two days. In the case of a rapidly rotating K1
subgiant, overlapping dynamo waves lead to noisy activity cycles 
and mixed-polarity fields at high latitudes.}
{The combined model reproduces the basic observed features of the solar 
cycle. 
The differences between the latitude distributions of the 
magnetic field at the bottom of the convection zone 
and the emerging surface flux grow with 
increasing rotation rate and convection zone depth, becoming
quite substantial for rapidly rotating dwarfs and subgiants.
The dynamical evolution of buoyantly rising magnetic flux 
should be considered as an essential ingredient in stellar dynamo models.}

\authorrunning{E. I\c{s}\i k et al.}
   \maketitle
%

\section{Introduction}
\label{sec:intro}
It is commonly accepted that the large-scale magnetic field of the Sun 
is maintained by a hydromagnetic dynamo process, which operates 
in the convective envelope and leads to an oscillatory magnetic field
with an average period of 11 years. 
Among the processes considered in the current theoretical context 
for the magnetic activity cycle
\citep[see reviews by][]{oss03, char05, fan09} are: 
(1) generation of toroidal magnetic field from poloidal field by 
    differential rotation ($\Omega$-effect), 
(2) formation and instabilities of toroidal magnetic flux tubes, 
(3) buoyant rise of toroidal magnetic flux, 
(4) generation of poloidal from toroidal magnetic field by 
helical convection 
    or through twisting of rising flux loops by the Coriolis force 
    ($\alpha$-effect), and 
(5) transport of magnetic flux by meridional flow.

In rapidly rotating stars with outer convection zones, manifestations 
of magnetic activity are observed throughout the electromagnetic 
spectrum. Distribution and coverage of starspots can be quite different 
from the solar activity patterns \citep{strass09}: spots
near the rotational poles, and high UV and X-ray fluxes 
indicative of a large surface coverage of magnetic regions. 
Observations of stellar magnetic activity provide 
constraints of stellar dynamo models \citep[see e.g.,][]{str05, msch05b, 
berdyugina05}. Hence, it is necessary to investigate the 
links between magnetic flux generation and transport, as well as the 
effects of stellar structure and rotation on the spatio-temporal 
distribution of the surface field. 

It is often tacitly assumed that the latitudinal distribution of toroidal 
magnetic fields generated in stellar interiors can be taken to represent 
the surface emergence patterns. 
However, the correspondence between the latitudinal distribution of the 
dynamo-generated field and the emergence pattern is not self-evident, 
because (a) the stability properties of flux tubes at the bottom of the 
convection zone depend sensitively on field strength and their position
in latitude \citep{afmsch93,afmsch95}, and 
(b) rising flux tubes in a rotating star are deflected towards the poles, 
owing to angular momentum conservation \citep[][]{chgil87,ss92,granzer00}.

Numerical simulations of the rise of flux tubes in the convection zone
\citep[e.g.,][]{dsilva93,fan94,cale95,cale98} have successfully
reproduced many observed properties of sunspot groups, among which the
tilt angle (the angle between the line connecting the opposite polarity
regions and the local latitudinal circle) is of particular importance
for Babcock-Leighton-type dynamos \citep[e.g.,][]{dikchar99}. Here, we
consider the connection between the dynamo mechanism operating in the
stellar interior, the transport of toroidal magnetic flux through the
convection zone, and the emerged flux evolving under the effects of
near-surface flows. We combine models for three processes: (1) a dynamo
operating in the overshoot layer at the bottom of the convection zone,
(2) the magnetic buoyancy instability and rise of magnetic flux tubes
through the convection zone, (3) the transport of magnetic flux at the
surface. In this first exploratory study we have used a very simple
one-dimensional dynamo model that yields the time-latitude
characteristics of the solar cycle. A more complete
two-dimensional dynamo model will be employed in a forthcoming study.

Preliminary results obtained with our combined model were presented
by \citet{isik07a}. In this paper, we give a detailed account of the
model setup, discuss the effects of different rotation rates in Sun-like
stars, and apply the model to a rapidly rotating K0-type main sequence
star, and a K1-type subgiant star.

\section{Dynamo model}
\label{sec:dynamo}

We consider the generation of magnetic flux in the overshoot layer at
the bottom of the convection zone (at radius $r_0 \simeq 0.73R_\odot$).
We assume a kinematic $\alpha\Omega$ dynamo operating in a thin layer
\citep{ss89}. The model is based on the assumption that the radial
diffusion of magnetic flux out of the layer is partly compensated by
downward flux pumping by convective flows. This thin-layer
$\alpha\Omega$ dynamo exhibits magnetic cycles and latitudinal propagation 
of dynamo waves based upon a radial gradient of the rotational angular
velocity.  Differential rotation in latitude and meridional circulation,
although probably relevant for the operation of solar/stellar dynamos
cannot be consistently incorporated into this simple model. We chose to
use this model as a simple means of obtaining solar-like time-latitude
behaviour of the toroidal field at the bottom of the convection zone and
to avoid overloading this exploratory study with the complications (and
uncertain parameters) of a more complex model. These models
can of course be incorporated into our model framework, as we intend to 
do in future studies.

\subsection{Dynamo equations}

We use spherical polar coordinates $(r,\theta,\phi)$ and decompose the 
azimuthally averaged magnetic field, ${\vec B}$, into a 
toroidal component, $(0,0,\hat{B}(r,\theta,t))$, and a poloidal component 
described by the vector potential, ${\vec A}=(0,0,\hat{A}(r,\theta,t))$, 
yielding 
\begin{eqnarray}
{\vec B} = \hat{B}(r,\theta,t){\vec e}_\phi + \vec\nabla\times \left( \hat{A}(r,\theta,t){\vec e}_\phi\right), 
\label{eq:decompose}
\end{eqnarray}
where ${\vec e}_\phi$ is the unit vector in the azimuthal direction. 
Following \citet{hoyng94}, the radial dependence of $\hat{A}$ and $\hat{B}$ 
is assumed to have the form of a spherical wave
\begin{eqnarray}
\hat{A}(r,\theta,t) &=& A(\theta,t)\frac{r_0}{r}\exp (ikr),
\nonumber \\
\hat{B}(r,\theta,t) &=& B(\theta,t)\frac{r_0}{r}\exp (ikr),
\label{eq:rdep}
\end{eqnarray}
reflecting flux loss from the thin dynamo
layer at $r=r_0$ by magnetic diffusion in the radial direction.
The mean magnetic field at $r=r_0$ as 
a function of the colatitude, $\theta$, and time, $t$, is
governed by the dynamo equations for $A(\theta,t)$ and $B(\theta,t)$, viz.
\begin{eqnarray}
\frac{\partial B}{\partial t} &=&
\Omega^\prime(\theta)\frac{\partial}{\partial\theta}(A\sin\theta) 
+\frac{B_0}{\tau}f\left(B\over B_0\right) \nonumber \\
& & +\frac{\eta}{r_0^2}\left[\frac{1}{\sin\theta}\frac{\partial}{\partial\theta}
\left(\sin\theta\frac{\partial B}{\partial\theta}\right)-\frac{B}{\sin^2\theta}
-(kr_0)^2 B \right],
\label{eq:dyn-d-t}
\\
\noalign{\vskip2mm}
\frac{\partial A}{\partial t} &=&
\alpha(\theta)B + 
\frac{\eta}{r_0^2}\left[\frac{1}{\sin\theta}\frac{\partial}{\partial\theta}
\left(\sin\theta\frac{\partial A}{\partial\theta}\right)-\frac{A}{\sin^2\theta}
-(kr_0)^2 A \right],
\label{eq:dyn-d-p}
\end{eqnarray}
where $\eta$ is the turbulent magnetic diffusivity,
$\Omega^\prime(\theta)$ is the radial gradient of the angular velocity
at $r=r_0$, and $\alpha(\theta)$ represents the $\alpha$-effect. For the
Sun, we take $r_0=5.07\times 10^{10}$~cm, which corresponds to the
middle of the convective overshoot layer 
(according to the stratification model used in Sect.~\ref{sec:emer}), 
and $kr_0=3$, so that a
quarter of the wavelength of the radial dependence of the magnetic field
corresponds roughly to the thickness of the convection zone.
The second term on the r.h.s. of Eq.~(\ref{eq:dyn-d-t}) represents the
buoyant loss of toroidal magnetic flux from the dynamo layer. The quantity 
$B_0$ is
the critical mean field strength, above which the flux is lost from the
layer with a characteristic timescale $\tau$, corresponding to the
growth time of the magnetic buoyancy instability of toroidal flux 
tubes (Sect.~\ref{sec:loss}).
The value of $B_0$ scales with the critical field strength for the
instability (average value between $0^\circ$ and $40^\circ$ latitude),
which depends on the given stellar model and rotation rate (see
Figs.~\ref{fig:sunall}a, \ref{fig:9dall}a, \ref{fig:2dall}a,
\ref{fig:k0vall}a, \ref{fig:k1all}a). 
The nonlinear function $f$ describes the $B$-dependence of the flux loss in the form
\begin{eqnarray}
f\left(\frac{B}{B_0}\right) = \left\{ \begin{array}{r@{\quad:\quad}l} -{\rm sgn}(B)
\cdot
\left({B}/{B_0}-1\right)^2 & {\rm if}~B \geq B_0 \\
                          0~~~~~~~~~~~~~~~~~~~~~~~ & {\rm otherwise},
               \end{array} \right.
\label{eq:b2}
\end{eqnarray}  
We rewrite the dynamo equations in nondimensional form by 
taking $r_0$ as our unit of length, the diffusion time, $r_0^2/\eta$, 
as the unit of time, $B_0$ as the unit field strength, $\alpha_0$ as the 
maximum absolute magnitude of the $\alpha$-effect, 
and $r_0 B_0$ as the unit of the vector potential. Using 
the same symbols as before, $(A,B)$, for the 
nondimensional quantities, Eqs.~(\ref{eq:dyn-d-t}) and (\ref{eq:dyn-d-p}) 
transform into 
\begin{eqnarray}
\frac{\partial B}{\partial t} &=&
R_\Omega\frac{\partial}{\partial\theta}(A\sin\theta)
+Q\cdot f(B)
 \nonumber \\ & & 
+ \frac{1}{\sin\theta}\frac{\partial}{\partial\theta} 
\left(\sin\theta\frac{\partial B}{\partial\theta}\right)
-\left(\frac{1}{\sin^2\theta}+(kr_0)^2 \right)B
\label{eq:dyn-nd-t}
\\
\noalign{\vskip2mm}
\frac{\partial A}{\partial t} &=&
R_\alpha B + 
\frac{1}{\sin\theta}\frac{\partial}{\partial\theta}
\left(\sin\theta\frac{\partial A}{\partial\theta}\right)-\left(\frac{1}{\sin^2\theta}
+(kr_0)^2 \right)A.
\label{eq:dyn-nd-p}
\end{eqnarray}
The induction effects are characterised by two dimensionless 
numbers, namely a Reynolds number for the $\alpha$-effect
\begin{eqnarray}
R_\alpha &=& \frac{\alpha_0 r_0}{\eta},
\label{eq:ra}
\end{eqnarray}
which represents the efficacy of induction by the $\alpha$-effect, 
and a Reynolds number for the radial shear ($\Omega$-effect) 
\begin{eqnarray}
R_\Omega &=& \frac{\Omega^\prime_0 r_0^3}{\eta},
\label{eq:ro}
\end{eqnarray}
which represents the efficacy of induction by the differential rotation.
The dimensionless number $Q$, which determines the strength of the
flux-loss nonlinearity is given by the ratio of the timescales for
diffusion and flux loss

\begin{eqnarray}
Q = \frac{\tau_d}{\tau} = \frac{r_0^2}{\eta\tau}.
\end{eqnarray}

In the dynamo model, we consider only one hemisphere and assume dipolar
parity for the magnetic field, i.e., antisymmetry with respect to
equatorial plane. The boundary conditions at the pole and at the equator
are thus given by
\begin{eqnarray}
B &=& A = 0~~{\rm for}~~\theta=0 \label{eq:bnd1} \\
\noalign{\vskip2mm}
B &=& 0,~~\frac{\partial A\sin\theta}{\partial\theta}=0~~{\rm for}~~\theta=\pi/2.
\label{eq:bnd2} 
\end{eqnarray}
We set the turbulent diffusivity to 
$\eta \simeq 2.96\times 10^{11}$ cm$^2$~s$^{-1}$. 
For the solar-type model, $\alpha_0 \simeq -11.7$~cm~s$^{-1}$ 
leads to a magnetic cycle period of 22 years.

Equations~(\ref{eq:dyn-nd-t})-(\ref{eq:dyn-nd-p}) with the boundary
conditions Eqs.~(\ref{eq:bnd1})-(\ref{eq:bnd2}) are numerically solved
with an implicit finite-difference method
\citep[cf.][]{ss89} using 90 grid points in latitude.


\subsection{Differential rotation and $\alpha$-effect}

The radial rotational shear in the dynamo layer,
$\Omega^\prime(\theta)=d\Omega(r,\theta)/dr\vert_{r=r_0}$, 
is taken in the form
\begin{eqnarray}
\Omega^\prime(\theta)=-\Omega_0^\prime P_2^0(\cos\theta)=
-\frac{\Omega_0^\prime}{2}(3\cos^2\theta+1), 
\label{eq:omega}
\end{eqnarray}
where $P_2^0(\cos\theta)$ is the associated Legendre polynomial.
Equation~(\ref{eq:omega}) with $\Omega^\prime_0 = 10^{-17}~{\rm
cm^{-1}~s^{-1}}$ roughly represents the shear profile in the solar
tachocline as determined by helioseismology \citep{schou98}.

\begin{figure}
\centering
\includegraphics[width=.8\linewidth]{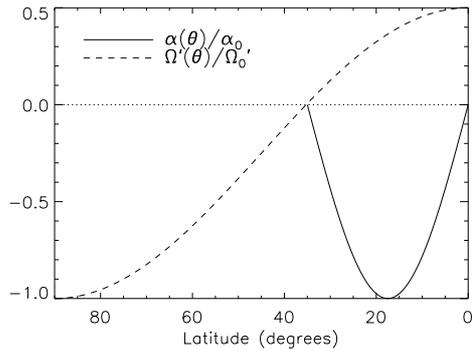}
\caption{Normalised latitudinal profiles of $\alpha$ 
and the radial shear $\Omega^\prime$. }
\label{fig:alpom}
\end{figure}

The latitudinal dependence of the $\alpha$-effect is assumed as 
\begin{eqnarray}
\alpha(\theta) = \left\{ \begin{array}{r@{\quad:\quad}l} 
\alpha_0\sin\left[ \pi \left(\theta-\theta_0\right)/\left(\pi/2-\theta_0\right)\right] & 
{\rm for}~\theta \geqslant \theta_0 \\
                              0~~~~~~~~~~~~~~~~~~~~~~~ & {\rm for}~\theta < \theta_0,
                              \end{array} \right.
\label{eq:alpha}
\end{eqnarray}
where $\alpha_0$ is the amplitude, and $\theta_0=55^\circ$, 
which corresponds to the zero crossing of the function 
$\Omega^\prime(\theta)$. The functions $\alpha(\theta)$ and 
$\Omega^\prime(\theta)$, normalised to their amplitudes, are shown in 
Fig.~\ref{fig:alpom}. In order for the dynamo waves 
to propagate equatorward, one must have $\alpha\Omega^\prime < 0$. 
To fulfil this criterion, $\alpha_0$ has been chosen to be negative, 
because the sign of radial shear in the solar tachocline is 
positive for $\theta>55^{\circ}$. 
The choice of Eq.~(\ref{eq:alpha}) and the sign of $\alpha_0$ are 
motivated by the $\alpha$-effect driven by magnetic flux tube 
instabilities \citep{afm94} and unstable magnetostrophic waves 
\citep{schmitt03}.

\section{Emergence of magnetic flux}
\label{sec:emer}

We assume that (1) the toroidal magnetic field in the dynamo layer
undergoes a magnetic Rayleigh-Taylor instability, which leads to the 
formation of magnetic flux tubes \citep{fan01,fan09}, 
(2) the flux tubes reach a 
mechanical equilibrium state by developing an internal flow 
\citep{mi92}, and (3) their equilibrium location is at the middle 
of the overshoot region. 
The flux tubes become subject to the undulatory (Parker) instability
once their field strength exceeds a critical value, $B_{\rm cr}$, which
is a function of latitude. We determine the linear stability properties
following \citet{afmsch95} and carry out numerical simulations of the
nonlinear evolution of the unstable flux tubes and their rise towards
the surface.

\subsection{MHD equations for magnetic flux tubes}
\label{ssec:basic}

The equation of motion for the material inside a flux tube, 
in a reference frame rotating with the angular velocity of the tube, 
${\vec \Omega}$, can be written as \citep{afmsch93} 
\begin{eqnarray}
    \rho_i\frac{D{\vec\varv}_i}{Dt} &=&
    -\vec\nabla\Big(p_i+\frac{B^2}{8\pi}\Big) + 
    \frac{({\vec B\cdot\vec\nabla}){\vec B}}{4\pi} \nonumber \\
    & & + \rho_i\big[{\vec g - \vec\Omega\times(\vec\Omega\times \vec r)}\big] + 
    2\rho_i{\vec \varv}_i\times\vec{\Omega} + {\vec F}_{\rm D},
\end{eqnarray}
where $D/Dt\equiv\partial/\partial t+{\vec\varv}\cdot\vec\nabla$ is the
Lagrangian derivative. The subscript $i$ denotes quantities inside the
flux tube and, in the following, the subscript $e$ denotes external
quantities. The terms on the right hand side of the equation are,
respectively, the total pressure force (gas and magnetic), magnetic
tension force, effective gravity (including the centrifugal force),
Coriolis force, and hydrodynamic drag force.  For the drag force, we use
the expression for a flow past a straight spherical cylinder,
\begin{eqnarray}
	{\vec F}_{\rm D} = -C_{\rm D}\frac{\rho_e \varv_\perp {\vec\varv}_\perp}{\pi R_{\rm t}},
	\label{eq:drag2}
\end{eqnarray}
where ${\vec\varv}_\perp=\varv_\perp {\vec {\hat{\varv}}}_\perp$ is the perpendicular 
component of the relative velocity of the tube with respect to the external medium, 
${\vec {\hat{\varv}}}_\perp$ is the unit vector along ${\vec\varv}_\perp$, $C_D$ is the 
hydrodynamic drag coefficient which we take as unity \citep{batchelor67}, 
$\rho_e$ is the density of the external medium, and $R_{\rm t}$ is the 
cross-sectional radius of the tube. 

The combination of the equations for continuity and magnetic induction 
in ideal MHD leads to Wal\'en's equation 
\begin{eqnarray}
\frac{D}{Dt}\left(\frac{{\vec B}}{\rho_i}\right) = 
\left( \frac{{\vec B}}{\rho_i} \cdot \vec\nabla \right){\vec\varv}_i. 
\label{eq:walen}
\end{eqnarray}
For the energy equation, we assume that the mass elements of the flux
tube evolve isentropically. This is justified because the timescale for
heat exchange by radiation in the deep convection zone is much longer
than the timescale of flux tube motions \citep{moreno86}.  The system
of equations is closed by the equation of state for an ideal gas and the
condition of magnetic flux conservation.

We consider the equations given above in the 
framework of the thin flux 
tube approximation \citep{spruit81}, in the form given by \citet{afmsch93}. 

\subsection{The escape of magnetic flux tubes from the dynamo layer}
\label{sec:loss}
Magnetic flux tubes leave the dynamo layer owing to the undulatory
buoyancy instability, which sets in for $B>B_{\rm cr}$ and leads to flux
loops rising through the convection zone. To determine the
number, field strength, and latitude of unstable flux tubes, we assign
a probability, $p$, for a flux loop to start rising at a given latitude
and time.

In the case of the Sun, we have $B_{\rm cr}\simeq 10^5$~G in the middle
of the convective overshoot layer \citep{msch94}. The
corresponding magnetic energy density is about 100 times higher than the
kinetic energy density of convective motions.  We assume that the
shear-generated mean toroidal field, $B$, is fragmented by the magnetic
Rayleigh-Taylor instability into flux tubes of strength $B_{\rm FT}$,
such that
\begin{eqnarray}
B = f B_{\rm FT},
\label{eq:fill}
\end{eqnarray}
where $f$ is the filling factor of flux tubes contributing to the mean
field.  For $B_{\rm FT}=10^5$~G and $f=0.1$, we have $B\simeq 10^4$~G,
which is of the order of the equipartition field strength, $B_{\rm
eq}=(4\pi\rho)^{1/2}\varv_c$, where $\varv_c$ is the convective velocity
determined by using a mixing length model. The individual flux tubes
are considered to be intensified to field strengths of the order of
$10^5$~G by a combination of rotational shear and the
conversion of potential energy in the external stratification 
as suggested by \citet{rempelsch01}.

\subsubsection*{Latitudes and number of erupting flux tubes}

\begin{figure}
\centering
\includegraphics[width=0.75\linewidth]{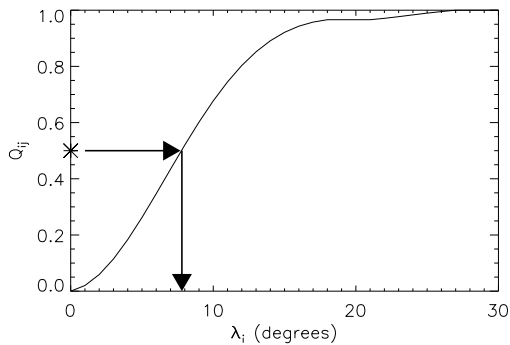}
\caption{A sample cumulative probability distribution function,
$Q_{ij}$, as a function of latitude, $\lambda_i$, at a given time
$t_j$. For each erupting flux tube, a pseudo-random number is chosen
in the interval [0,1]. The
latitude $\lambda_i$ at which the flux tube erupts is then obtained by
the mapping shown by the arrows. }
\label{fig:sampleq1}
\end{figure}

We assume the emergence rate of flux tubes to be proportional to the 
mean toroidal field given by the dynamo model,
\begin{eqnarray}
B_{ij} =B(\lambda_i,t_j)
\end{eqnarray}
for a given latitude, $\lambda_i$, and time, $t_j$, which are discretised 
into time intervals, $\Delta t=t_{j+1}-t_j$, and latitude intervals, 
$\Delta\lambda=\lambda_{i+1}-\lambda_i$, with $\lambda=\pi/2 - \theta$ 
and $-\pi/2\leqslant\lambda\leqslant\pi/2$. 
The probability of a flux tube at $\lambda_i$ and $t_j$ to start rising 
is assumed to be proportional to the local mean field, so that we have
\begin{eqnarray}
p_{ij} = \frac{B_{ij}}{\sum_i B_{ij}}.
\end{eqnarray}
We denote the number of flux tubes emerging 
at time $t_j$ with $n_j$. To determine the initial 
latitudes of flux tubes at the bottom of the convection zone, 
$n_j$ pseudo-random numbers taken from a uniform distribution in the 
interval $[0,1]$ are mapped via the cumulative probability 
distribution function (PDF), $Q_{ij}$, based on $p_{ij}$: 
\begin{eqnarray}
Q_{ij} = \sum_{k=0}^i p_{kj}.
\end{eqnarray}
The mapping for a sample PDF is illustrated in Fig.~\ref{fig:sampleq1}. 
%
The number of emerging tubes, $n_j$, is chosen to be proportional to 
the total toroidal flux density at that time, $\sum_i B_{ij}$, requiring 
that a total number of $N$ tubes emerge per activity cycle, viz.
\begin{eqnarray}
n_j= \left[N\cdot\frac{\sum_i B_{ij}}{\sum_{i,j} B_{ij}}+0.5\right].
\label{eq:nofts}
\end{eqnarray}
Here the brackets denote the nearest integer to the value of the 
expression they enclose. 
For the case of the Sun, $N$ is set to 2100, which constrains the 
total flux emerging over an activity cycle to be of the order 
of $10^{25}$~Mx, which is comparable with observations. 
\begin{figure}
\centering
\includegraphics[width=0.75\linewidth]{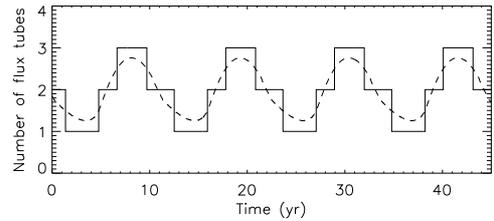}
\caption{Number of flux tubes per time interval of 7 days, $n_j$, as a
function of time. The dashed line shows the value in brackets of
Eq.~(\ref{eq:nofts}), and the full line shows $n_j$.}
\label{fig:nofts2}
\end{figure}
Figure~\ref{fig:nofts2} shows $n_j$ as a function of time, with $B(\lambda,t)$ 
taken from the solar dynamo model given in Sect.~\ref{sec:dynamo}. 

\subsubsection*{Field strengths of unstable flux tubes}

\begin{figure}
\centering
\includegraphics[width=0.5\linewidth]{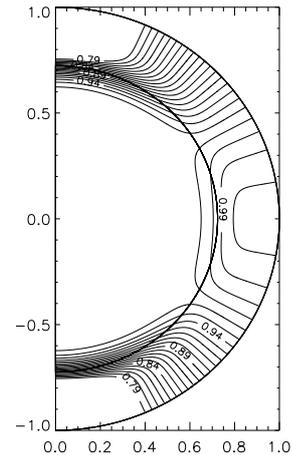}
\caption{Meridional profile of internal rotation used in 
the stability analysis and flux tube simulations. 
Contours denote angular velocity, normalised to the equatorial 
value at the surface. }
\label{fig:introt}
\end{figure}

To determine the field strength $B_{\rm FT}$, 
we consider the stability properties of flux tubes \citep{afmsch95} 
in the middle of the overshoot region (5000 km above the upper 
boundary of the radiative zone), using the non-local mixing length model 
of \citet{skastix91}. 
We consider a rotation profile similar to that 
obtained using helioseismic data \citep[e.g.,][]{schou98} 
\begin{eqnarray}
\frac{\Omega(r,\theta)}{\Omega_0} &=& 
0.9635 - \left[1+{\rm erf}\left(\frac{r-r_0}{d_0}\right)\right]
\nonumber \\
& & \cdot\left(0.0876\cos^4\theta+0.0535\cos^2\theta-0.0182\right),
\label{eq:introt}
\end{eqnarray}
where $\Omega_0$ is the equatorial rotation rate at the surface, 
and $d_0=0.075R_{\sun}$ is taken as the half-thickness of the tachocline. 
Lines of constant angular velocity are shown in Fig.~\ref{fig:introt}. 
This 2D profile is consistent with the latitude dependence of the radial 
shear given in Eq.~(\ref{eq:omega}), which we use in the dynamo model. 
%
%

Figure~\ref{fig:sunall}a shows the stability diagram in a plane defined by 
flux tube latitude and field strength. 
We define $B_{\tau}(\lambda)$ to be the field strength of unstable 
flux tubes as a function of latitude, corresponding to a certain growth 
time, $\tau$, of the instability. 
In the course of the amplification of the toroidal magnetic field,
instability sets in when $B_{\rm FT}$ exceeds a critical value. 
This value also defines the mean field strength $B_0$ in
Eq.~(\ref{eq:b2}), at which the dynamo saturates, taking into account
the scaling given by Eq.~(\ref{eq:fill}).
In most parts of the stability diagram, $\tau$ decreases with increasing
field strength.
In the $\alpha\Omega$ dynamo model, the toroidal field amplification
occurs on the timescale of the rotational shear in the tachocline,
which is about a few years.
As long as the growth time of the flux tube instability is longer than
the shearing timescale, the toroidal field continues to be amplified
without the flux tube leaving the shear region \citep{isik09}.
As the field strength reaches a level of the order $10^5$~G, the two
timescales become comparable and the flux tube is lost from the layer.
Therefore, we do not consider unstable tubes with growth times much
longer than the amplification timescale and choose the field strengths
of the emerging flux tubes given by the curve $B_{200}(\lambda)$,
corresponding to $\tau=200$ days.  Choosing a growth time exceeding
this value leads to almost identical results since the curves of constant
growth time become very close to each other near the stability limit
(Fig.~\ref{fig:sunall}a).  We consider the contour line
$B_{200}(\lambda)$ of the main region of instability in
Fig.~\ref{fig:sunall}a, which is a single-valued function of latitude.
Taking field strengths corresponding to the low-latitude ``island'' to
the left of the diagram does not lead to a significant difference in the
emergence latitudes and tilt angles \citep{cale95}.
To avoid the occasional emergence of flux loops at
unrealistically high latitudes in the case of the Sun, we assume that
unstable flux tubes only form if the local mean toroidal field exceeds
400~G (4\% of the equipartition field strength). 



\subsection{The rise of flux loops}
\label{sec:rise}

Having determined the times of emergence, initial latitudes, and the
corresponding field strengths of unstable flux tubes, we simulate their
rise through the convection zone using the code
developed by \citet{moreno86} and extended to three dimensions and
spherical geometry by \citet{cale95}. In the numerical scheme, the thin
flux tube is described by a string of Lagrangian mass elements.

The initial value for the cross-sectional radius of each flux tube is
taken to be $R_{\rm t}=1000$~km.  For $B=10^5$~G, this corresponds to a
magnetic flux of about $3\times 10^{21}$~Mx, which is typical of a
bipolar magnetic region (hereafter BMR) of moderate size at the solar
surface.  The tube radius is large enough for the rising flux loops to
be largely unaffected by the drag force \citep{dsilva93, cale95}, 
so that differential rotation and meridional circulation do not
influence their motion. Therefore, the dynamics of flux tubes with
$R_{\rm t}\gtrsim 1000$~km is independent of their radius, as long as
the thin flux tube approximation remains valid.

The fastest growing modes of unstable flux tubes have azimuthal
wavenumbers $m=1$ or $m=2$. On the other hand, the angular separation of
large BMRs on the Sun corresponds to much larger azimuthal wavenumbers
($m=10-60$). A possible mechanism for limiting the longitudinal
separation of BMRs is the dynamical disconnection from their magnetic
roots at depths of a few Mm below the surface soon after emergence
\citep{sch+rmp05}.

To determine the emergence latitudes and tilt angles, we use a
table derived from a number of simulations of rising unstable flux tubes
for various initial latitudes and magnetic field strengths.  The initial
latitudes are in steps of $5^\circ$ and the corresponding field
strengths are obtained from the curve $B_{200}(\lambda)$. Linear 
interpolation is used to obtain the emergence latitudes and tilt angles
at intermediate values of initial latitude and field strength.
Figure~\ref{fig:sunall}b shows a comparison of the dynamo-generated
toroidal field in the overshoot layer (contour lines) and the surface
emergence pattern of rising flux loops (dots).
The overall emergence pattern coincides with the dynamo waves, since the
poleward deflection of rising flux tubes with field strength of the
order of $10^5$~G by the Coriolis force is small for solar rotation
rates.  This result is consistent with the implicit assumption often
made when interpreting solar dynamo models, namely that the surface
activity pattern closely reflects the dynamo wave pattern.

\section{Surface flux transport}
\label{sec:sft}

Times, latitudes, and tilt angles of emerging flux loops
obtained with the procedures described so far determine the flux input
into the surface flux transport model, which we consider in this
section. The model describes the 
subsequent evolution of magnetic flux at the stellar surface.

\subsection{The surface evolution of magnetic flux}
\label{sec:evo}

We follow the evolution of the surface field governed by the emergence
of BMRs and flux transport by differential rotation, meridional flow,
and {(turbulent) diffusion due to convective flows. In contrast to the
big flux tubes rising through the convection zone, the large-scale
surface field described by the flux transport model is highly fragmented
into small-scale flux concentrations. These magnetic elements are
vertically oriented (owing to buoyancy) and sufficiently small to be
passively transported by the surface flows via the drag force.}

We therefore assume that the surface field has only a radial component,
$B_r$, whose evolution as a function of latitude, $\lambda$, longitude,
$\phi$, and time, $t$, is described by the magnetic induction equation
in the form \citep[cf.][]{bsss04}
\begin{eqnarray} 
\frac{\partial B_r}{\partial t} &=& 
- \Omega(\lambda) \frac{\partial B_r}{\partial \phi}      
+ \frac{1}{R_{\star}\cos\lambda}\frac{\partial}{\partial\lambda} 
\Big(\varv(\lambda) B_r\cos\lambda \Big) \nonumber \\ \noalign{\vskip2mm}
& & + \mathcal{D}_h(\eta_h) 
+ \mathcal{D}_r(\eta_r) + \mathcal{S}(\lambda,\phi,t),
\label{eq:transport}
\end{eqnarray}
where $R_\star$ is the stellar radius, $\mathcal{D}_h$ is the term for
horizontal diffusion with uniform turbulent diffusivity
$\eta_h=600$~km~s$^{-1}$, $\mathcal{D}_r$ denotes the radial diffusion
term with uniform radial diffusivity $\eta_r=100$~km~s$^{-1}$
\citep{bss06}. In addition, $\mathcal{S}$ is the source term describing the newly
emerging BMRs and $\Omega$ is the surface angular rotation rate as a
function of latitude $\lambda$
\begin{eqnarray}
\Omega(\lambda) = 13.38 - 2.30\sin^2\lambda - 1.62\sin^4\lambda \;\;\;\;
\mathrm{degrees~day^{-1}} \label{eq:dr}
\end{eqnarray}
\citep{snod83}. The meridional flow velocity $\varv$ 
\citep[][]{snodd96, hath96} is assumed to be
\begin{eqnarray}
\varv(\lambda) = \left\{ \begin{array}{r@{\quad:\quad}l}-\varv_0\sin(\pi\lambda / \lambda_\mathrm{c}) & {\rm if}~|\lambda| < \lambda_\mathrm{c} \\
                              0 & {\rm otherwise},
                              \end{array} \right.
\label{eq:mf}
\end{eqnarray}
where $\varv_0=11~\mathrm{m~s^{-1}}$ and $\lambda_\mathrm{c}=\pm~75^{\circ}$
\citep[cf.][]{balle98, bsss04}.

The numerical solution of Eq.~(\ref{eq:transport}) is carried out by
representing the magnetic field as an expansion in terms of spherical
harmonics with a maximum degree of $\ell=63$ \citep{bsss04}. This
corresponds approximately to the observed size of the supergranules
($\sim 30$~Mm) on the Sun.  

\begin{figure*}
\centering
\includegraphics[width=0.45\linewidth, bb=28 28 532 388]{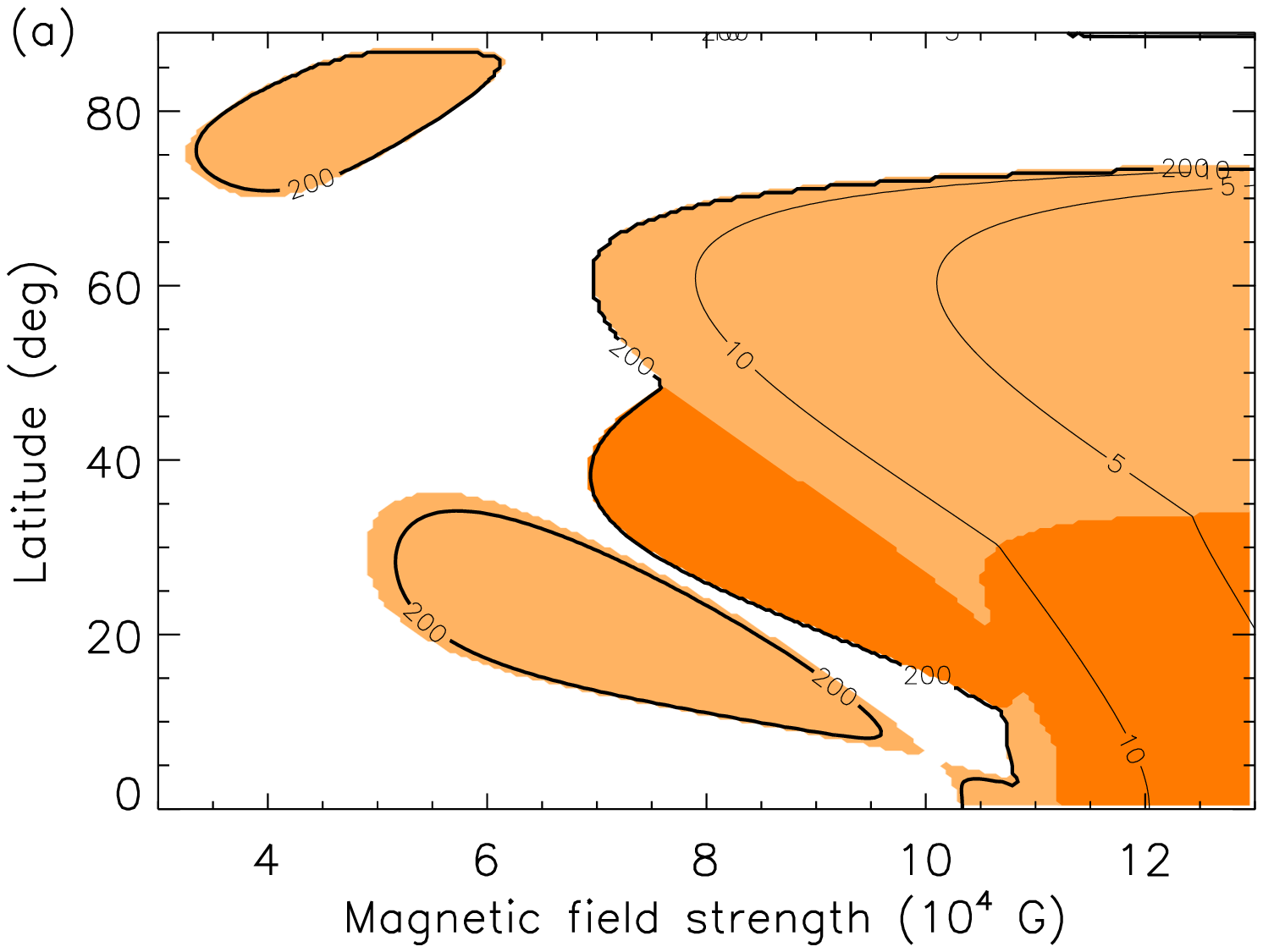}
\includegraphics[width=.5\linewidth]{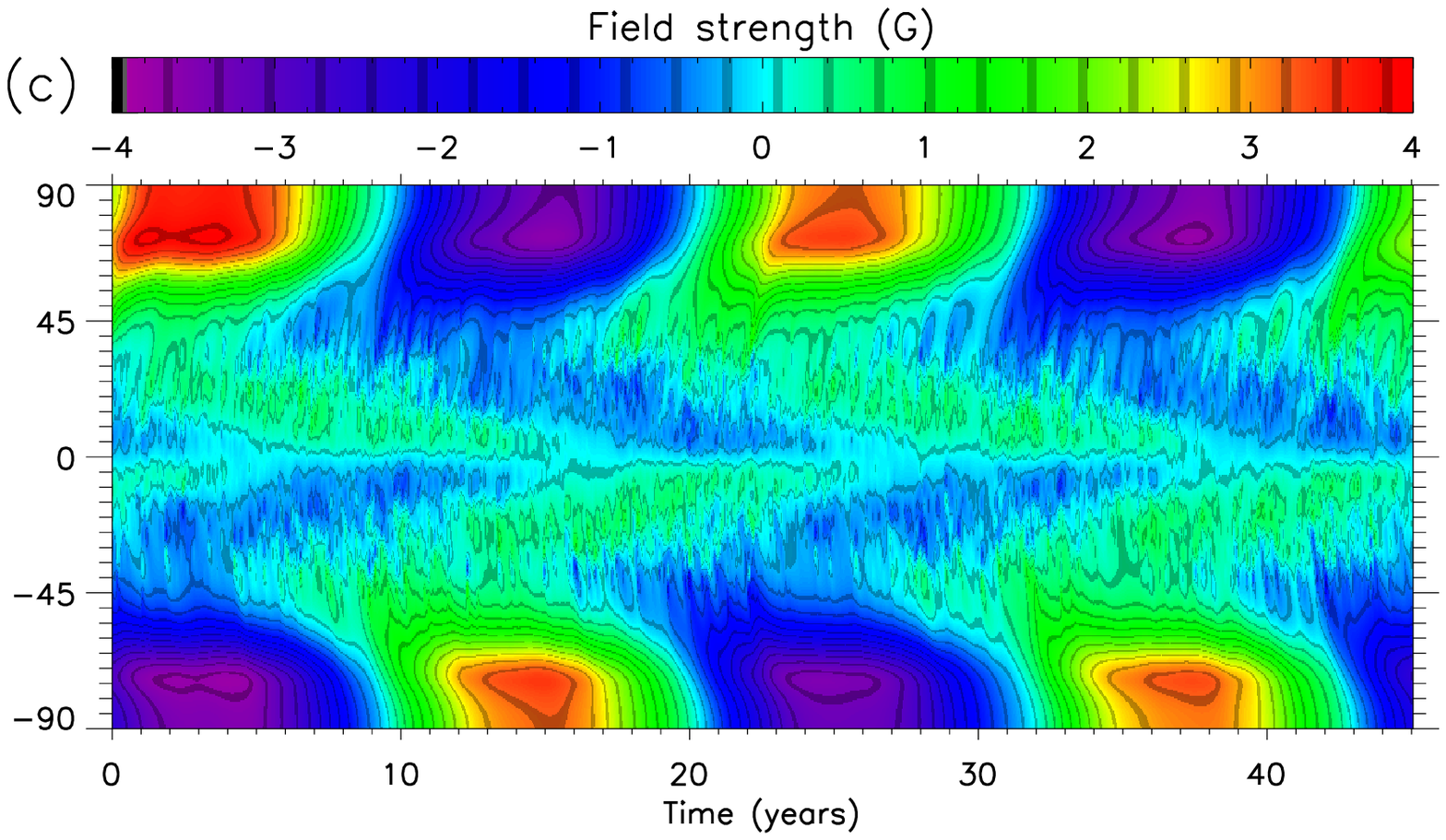}\\
\includegraphics[width=.45\linewidth]{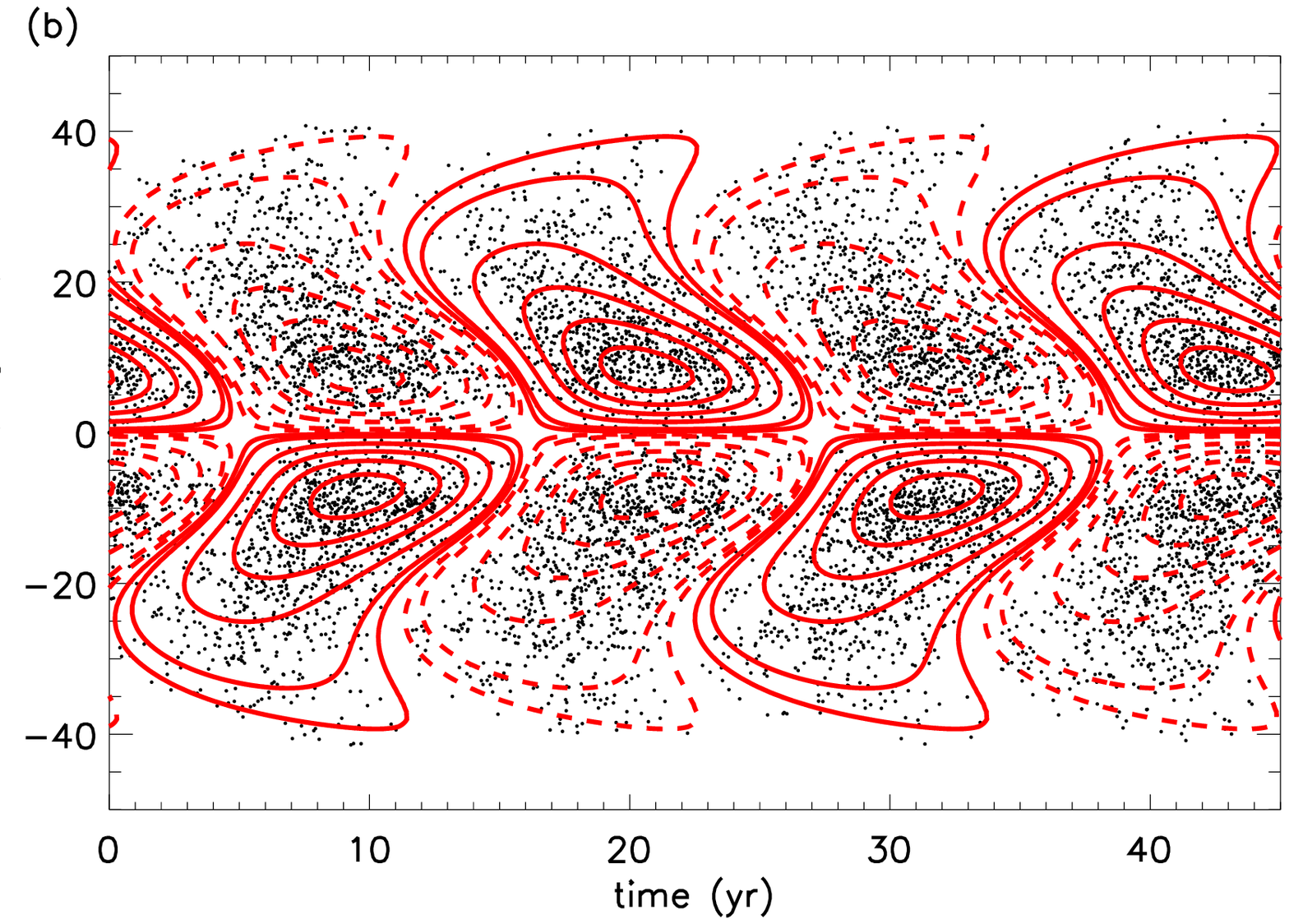}
\includegraphics[width=.5\linewidth]{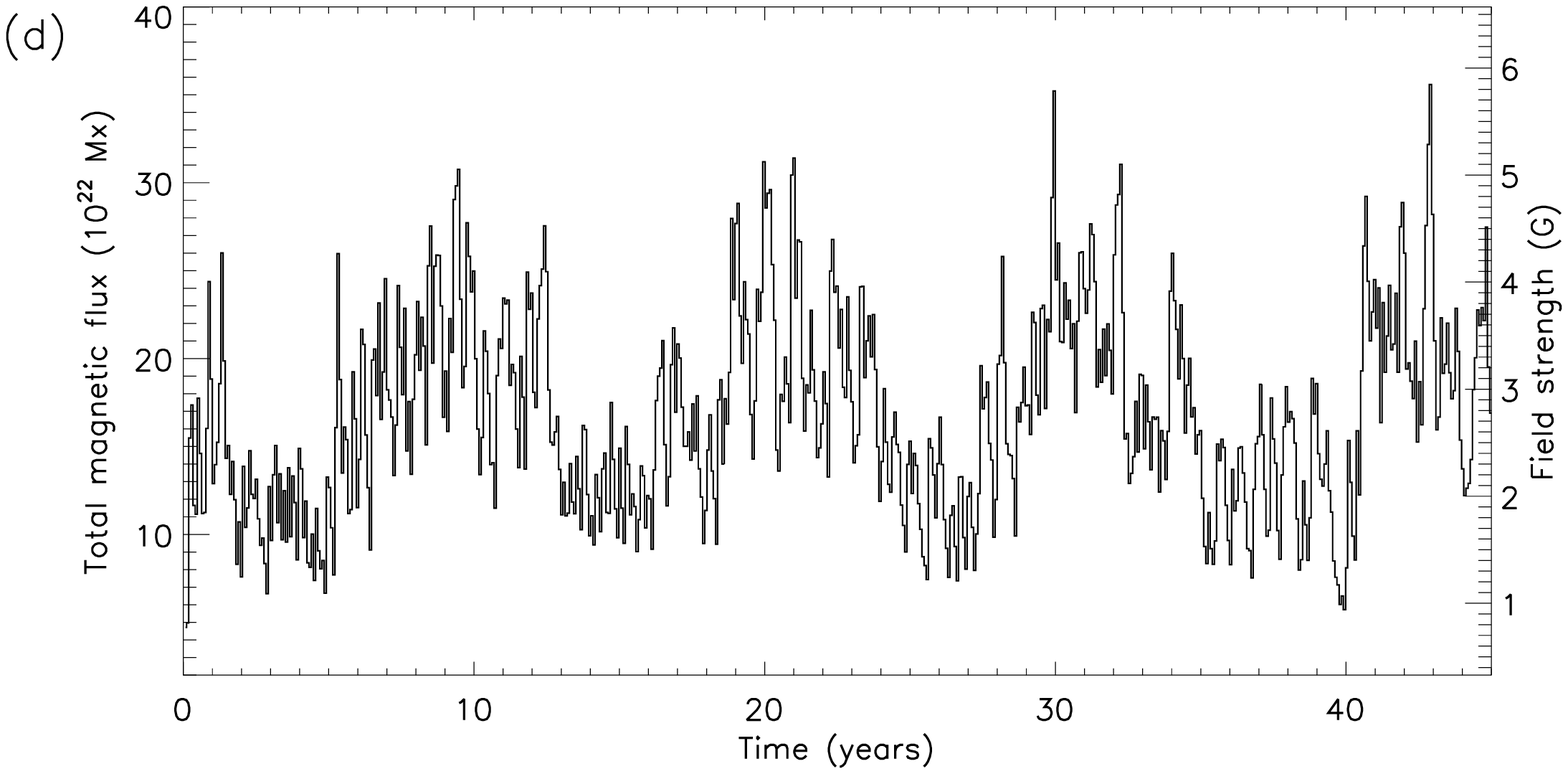}
\caption{(a) Stability diagram for thin magnetic flux tubes in the middle of the 
solar overshoot region, for Sun-like differential rotation (given by 
Eq.~\ref{eq:introt}). White 
regions represent stable flux tubes, while shaded areas indicate instability. 
For light-shaded areas, the fastest growing mode has azimuthal wave number 
$m=1$, while for dark-shaded areas it is $m=2$. Contours show lines of 
constant growth times (e-folding times), labelled in units of days. 
(b) Time-latitude diagram of the dynamo-generated mean toroidal
   magnetic field $B(\lambda,t)$ at the bottom of the convection zone
   (contours at $\pm0.04$, $\pm0.1$, $\pm0.3$, $\pm0.5$, $\pm0.7$, $\pm0.9$ 
   of the maximum value of $B$, solid for positive and dashed for
   negative values) and of the flux loops emerging at the surface (dots) 
   for a Sun-like star. 
(c) Time-latitude diagram of the longitudinally averaged
   radial surface magnetic field for the solar model.  
(d) Time variation of the total unsigned surface magnetic flux. The values 
   are averaged over 27-day time intervals.}
\label{fig:sunall}
\end{figure*}


%
\begin{figure}
\includegraphics[width=\linewidth]{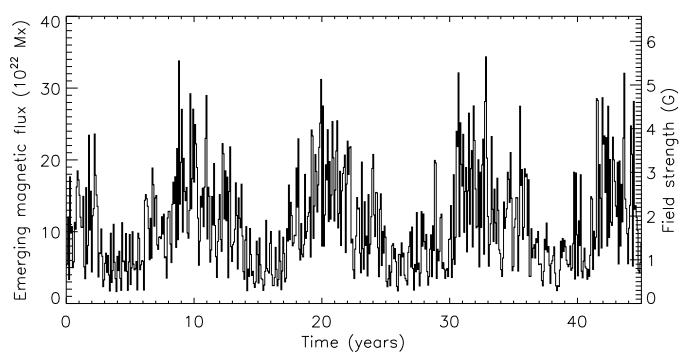}
\caption{Time variation of the emerging magnetic flux integrated over 
the surface, and summed over 27-day time intervals. The corresponding 
average unsigned field strength is indicated on the right-hand vertical 
axis.}
\label{fig:arfluxp27}
\end{figure}

\subsection{Treatment of source BMRs}
\label{sec:source}

The flux density of a newly emerged BMR is written in the form
\begin{eqnarray}
      \label{eq:net}
      B_r(\lambda,\phi) = B^+(\lambda,\phi) - B^-(\lambda,\phi).
\end{eqnarray}
The emergence longitudes of the BMRs are randomly distributed.
Following \citet{balle98} and \citet{bsss04}, we assume that the unsigned
field strength of the two polarities is given by
\begin{eqnarray}
      \label{eq:gaudist}
      B^{\pm}(\lambda,\phi) = B_0 \exp\left[- \frac{ 2\left[ 1-\cos\beta_{\pm}(\lambda,\phi) \right]}{\delta_{\rm in}^2}\right].
\end{eqnarray}
where $ \beta_{\pm}(\lambda,\phi) $ are the heliocentric angles between a
given position $(\lambda,\phi)$ and the centres of the positive and negative
polarities, $(\lambda_{\pm},\phi_{\pm})$. 
The initial characteristic angular width of each polarity, 
$\delta_{\rm in}$, and 
the angular separation of the centres of the two poles, $\Delta\beta$, are 
related by $\delta_{\rm in}=0.4\Delta\beta$. 
The maximum field strength $B_0$ is set to 250 G \citep[see][]{bsss04}. 

The BMRs, which are smaller than the grid cell size ($\sim 1$ square degree)
of the simulations, are taken into account by considering these regions
only at a later stage in their evolution, after having expanded (by
turbulent diffusion) to an angular width of $\delta_0=4^\circ$.  The
description of the two polarities for such a BMR is thus given by
\begin{eqnarray}
      B^{\pm}(\lambda,\phi) = B_0 \Bigg(\frac{\delta_{\rm
in}}{\delta_0}\Bigg)^2\exp\left[- \frac{ 2\left[
1-\cos\beta_{\pm}(\lambda,\phi) \right]}{\delta_0^2}\right].
\end{eqnarray}
In our numerical treatment, the spherical harmonic coefficients defining
the field distribution of the newly emerged BMRs are multiplied by a
spatial filter of the form $\mathrm{exp}\big[-\beta_\pm^2 l(l+1)/4\big]$
\citep{balle98}, in order to diminish the effect
of ringing (Gibbs phenomenon) caused by the truncation of the expansion
in spherical harmonics.

\subsection{Size distribution and the emerging magnetic flux}
\label{ssec:size}

The total size of a newly emerged BMR is related to the heliocentric
angular separation of the centres of the two poles, $\Delta\beta$,
which, in our model, ranges from $3.5^\circ$ to $10^\circ$ in steps of
$0.1^\circ$.  The corresponding range of areas is 30$-$250 square
degrees, and the resulting range of magnetic flux is
$2.3\times10^{21}-3.5\times10^{22}$ Mx. The areas $A$ of the bipolar
source regions are determined randomly with a number distribution
derived from solar observations, $N(A)\sim A^{-2}$ \citep{schrharvey94}.

For the solar model, the time series of emerging magnetic flux, summed
over 27 days (one full solar rotation), is given in
Fig.~\ref{fig:arfluxp27}.
It shows a cyclic variation with a period of about 11 years, closely
following the underlying oscillatory dynamo model. 

The evolution of the surface flux is given in a supplementary 
animation (see the online appendix, Fig.~\ref{fig:bsurf}, top panels), 
which shows the weakening and cancellation of the polar field, and 
the gradual build-up of the polar field of the following cycle. 
The time-latitude diagram of the longitudinal average of the signed flux
density is shown in Fig.~\ref{fig:sunall}c.  The extreme values of the
unsigned flux density are lower than those found by \citet{bsss04} for a
similar model. The reason is that the tilt angles resulting from the
simulated rise of flux tubes are systematically smaller than the
latitude dependence assumed by \citet{bsss04}, but actually provide a
better match to the observations \citep[see][Fig.~12]{cale95}.  The
corresponding time variation in the total unsigned flux (and the
equivalent average unsigned field strength) is shown in
Fig.~\ref{fig:sunall}d.  The activity minima are less marked than those of
the emerging flux (Fig.~\ref{fig:arfluxp27}); this is due to the phase
difference of about half a period between the variations in the polar
flux and the flux emerging at low latitudes.

\section{Rapidly rotating Sun-like stars}

As a first non-solar application of the combined model of magnetic flux
generation and transport, we consider rapidly rotating stars of solar
internal structure.

For simplicity (and due to lack of information), 
we assume the same values for
the turbulent magnetic diffusivities for the dynamo region, the
convection zone, and the surface as in the solar model discussed in the
previous sections. We also use the same profiles for the rotational
shear, i.e., we assume that the difference between the minimum and
maximum angular velocities in the convective envelope, $\Delta\Omega$,
is independent of the equatorial rotation rate at the
surface, $\Omega_{\star}$. This is motivated by observations of cool
stars indicating a rather weak dependence of $\Delta\Omega$ at the
stellar surface on $\Omega_{\star}$
\citep[e.g.,][]{barnes05,donahue96}. This assumption corresponds to a
decrease in $\Delta\Omega/\Omega_{\star}$ by a factor
$\Omega_{\odot}/\Omega_{\star}$ with respect to the solar value, so that
the constants on the right hand side of Eq.~(\ref{eq:introt}) have to be
multiplied by $\Omega_{\odot}/\Omega_{\star}$.  Because $\Delta\Omega$
is kept constant, the surface flux transport equation 
(Eq.~\ref{eq:transport}), the rotation profile (Eq.~\ref{eq:dr}), 
and $R_\Omega$ (Eq.~\ref{eq:ro}) remain unchanged. We also adopt a 
solar-type internal rotation profile in the present section, owing to 
the lack of observational information on stellar internal rotation profiles. 
When results from asteroseismology become available, they can be directly 
incorporated into the model.

In the dynamo model, we assume that $\alpha_0$ scales with
$\Omega_{\star}$ \citep{krause80}, so that $R_\alpha\propto\Omega_\star$
in Eq.~(\ref{eq:ra}), because $\eta$ is not changed.  The dynamo cycle
period, $P_{\rm cyc}$, then decreases with increasing
$\Omega_{\star}$. We note, however, that this scaling might not be valid
for the very most rapid rotators since the $\alpha$-effect driven by
helical turbulence saturates when the cyclonic rotation of rising fluid
parcels exceeds $\pi/2$ \citep{parker82,rk93}.

The total number of erupting flux tubes per activity cycle, $N$, is also
scaled with $\Omega_{\star}$, in accordance with the observed relation
between the rotation rate and the level of magnetic activity
\citep[e.g.,][]{montesinos01}.

\subsection{The case of $P_{\rm rot}=9~{\rm d}$}
\label{sec:10d}

\begin{figure*}
\centering
\begin{minipage}{\linewidth}
\includegraphics[width=0.45\linewidth, bb=28 28 532 388]{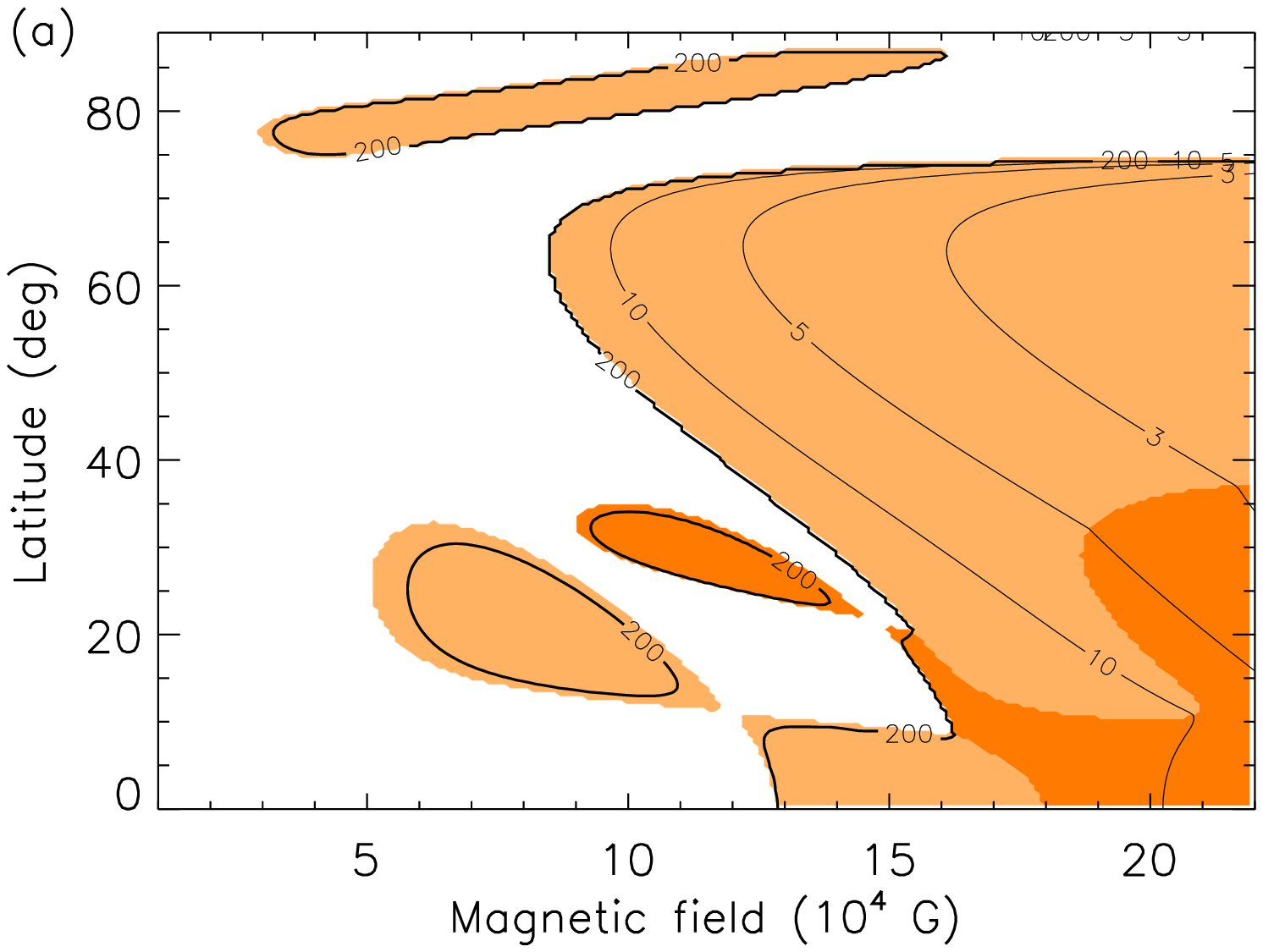}
\includegraphics[width=.5\linewidth]{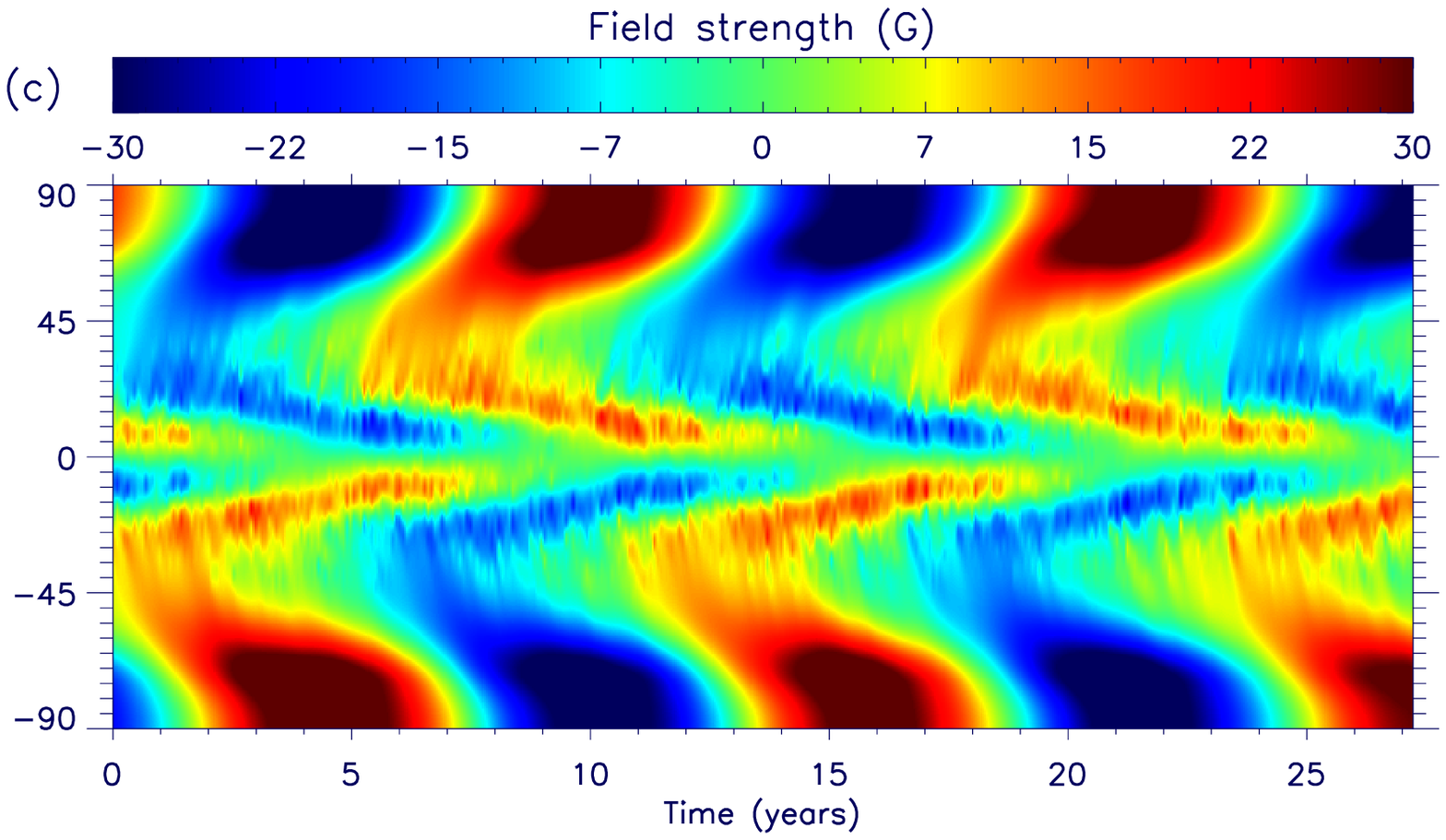} \\ 
\includegraphics[width=.45\linewidth]{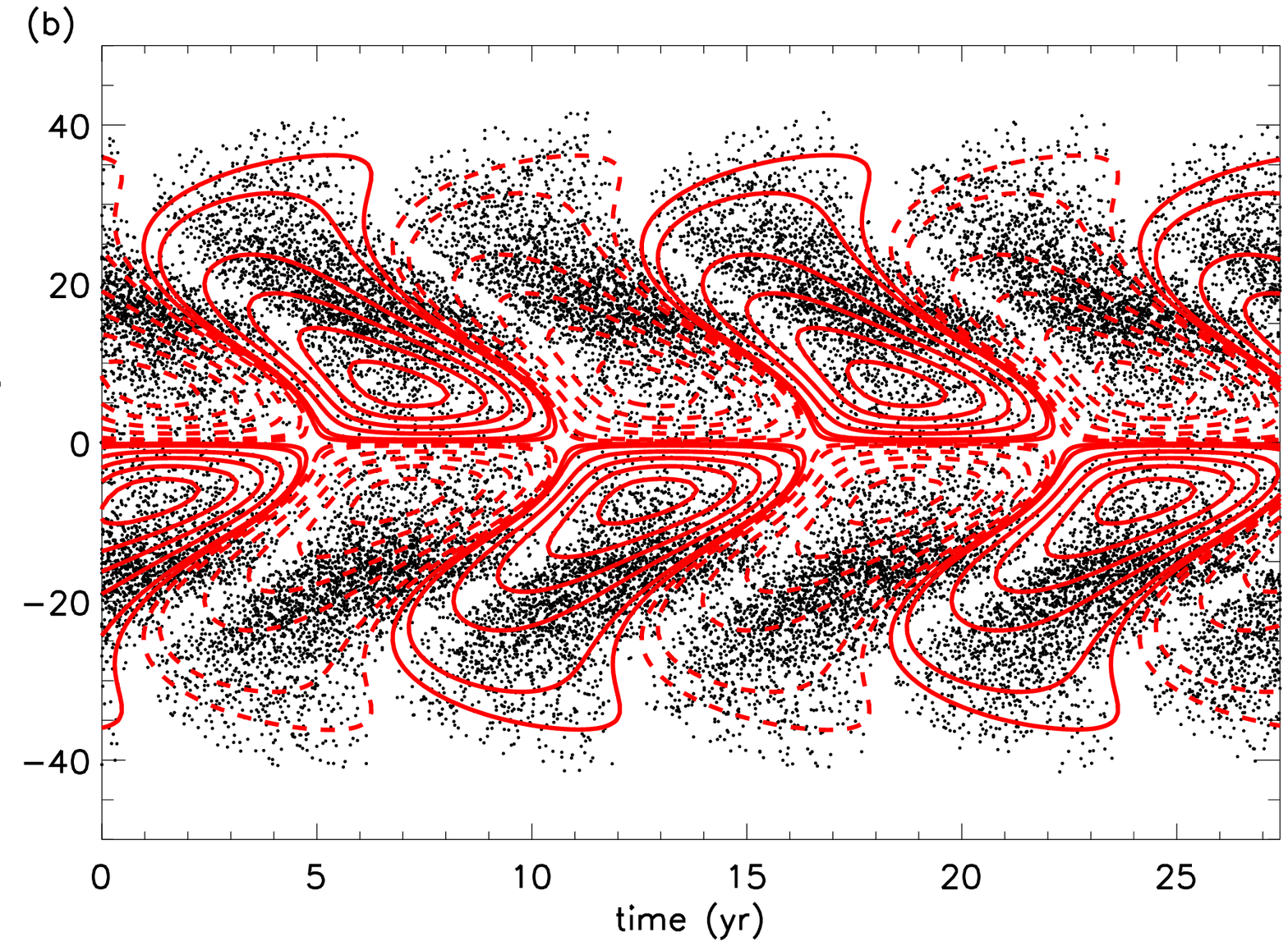} 
\includegraphics[width=.5\linewidth]{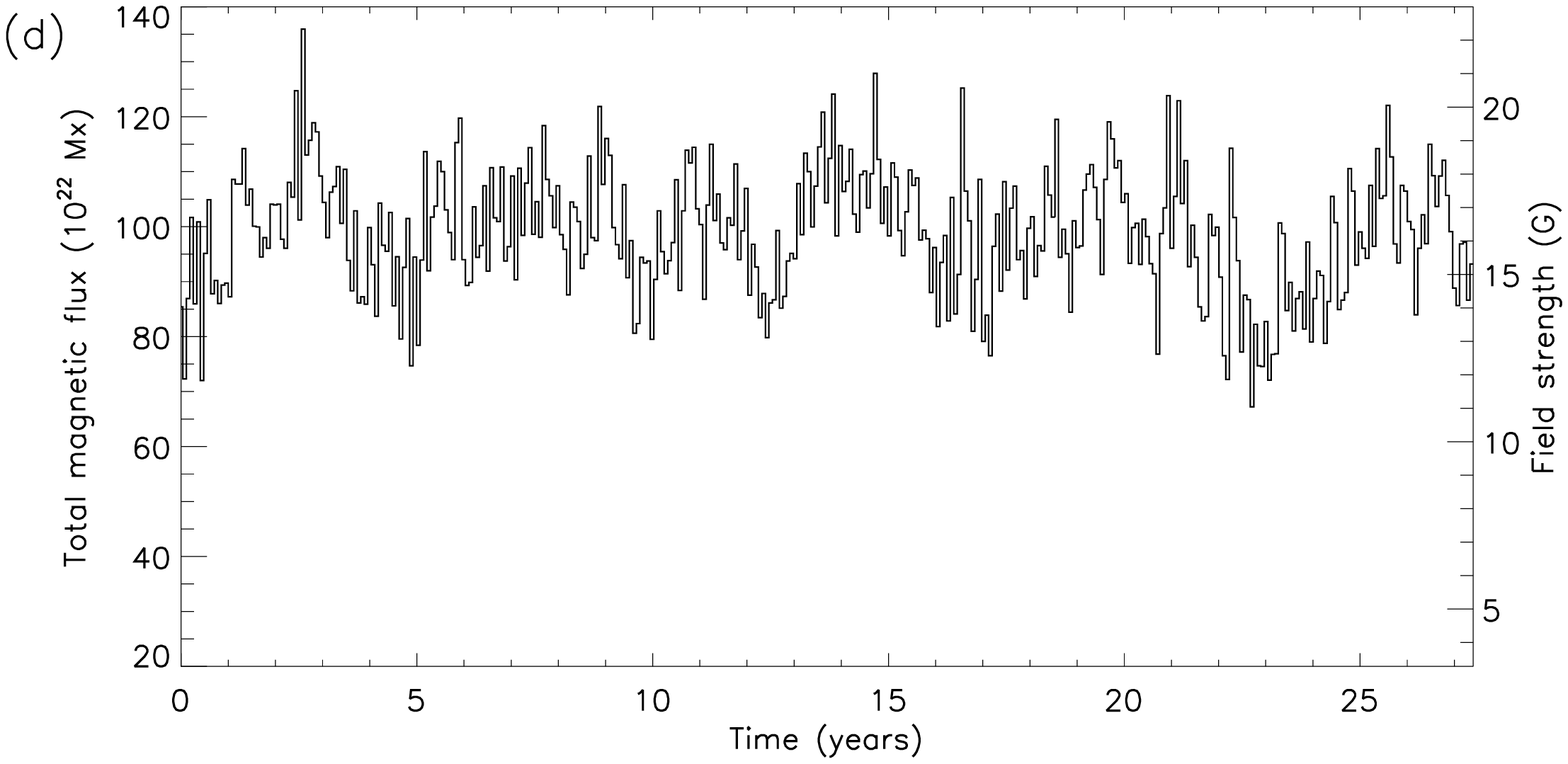}
\end{minipage}
\caption{Same as Fig.~\ref{fig:sunall}, but for a Sun-like star with 
$P_{\rm rot}=9$~d.}
\label{fig:9dall}
\end{figure*}

We first consider a Sun-like star with an equatorial rotation rate of
$\Omega_\star=2.7\Omega_\odot$, corresponding to a period of about
9~d. 
Figure~\ref{fig:9dall}a
shows the stability diagram of flux tubes located in the middle of the
overshoot region. 


The critical field strength for the Parker instability is larger than in the
solar case, since rotation has a stabilising effect on the flux
tubes. With increasing $~\Omega_\star$, angular momentum conservation
suppresses the tube displacements perpendicular to the rotation axis, so
that the magnetic buoyancy needs to be stronger than in a similar
situation in the Sun.

In determining the function $B_{200}(\lambda)$, we excluded the
`islands' of instability seen in Fig.~\ref{fig:9dall}a. Test simulations
showed that the emergence latitudes and tilt angles of rising flux tubes
do not differ significantly between the islands and the neighbouring part
of the main region of instability.

The time-latitude diagram of the dynamo-generated toroidal field at the
bottom of the convection zone and the locations of flux emergence are
combined in Fig.~\ref{fig:9dall}b.  They show a cyclic variation with
the dynamo period of about five years. The poleward deflection of rising
flux tubes is slightly stronger than in the case of solar rotation. The
effect responsible is the component of the Coriolis acceleration 
directed
towards the rotation axis (resulting from a retrograde flow along the
flux tube required by angular momentum conservation), which is proportional
to the rotation rate.

\begin{figure}
\includegraphics[width=\linewidth]{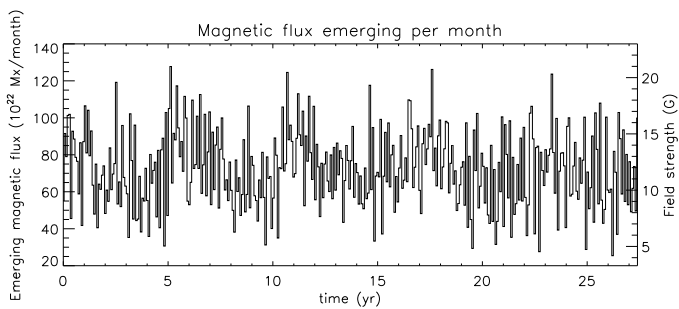}
\caption{Same as Fig.~\ref{fig:arfluxp27}, but for a Sun-like star with 
$P_{\rm rot}=9$~d.}
\label{fig:arfluxp10}
\end{figure}

The emerging magnetic flux, summed over one-month intervals 
is shown in Fig.~\ref{fig:arfluxp10}. 
While the total flux emerging per cycle is about five times higher than in
the solar case, the relative variation between the minima and maxima of 
the emerging flux is considerably weaker.  This is due to the larger
overlap between consecutive cycles, which is a consequence of stronger
dynamo excitation. Figure~\ref{fig:9dall}c shows that also the polar field
maxima are higher than in the case of solar rotation (note the different
saturation level of the colour table from
Fig.~\ref{fig:sunall}c). This is due to 1) a higher flux emergence rate and
2) the bigger tilt angles of the emerging BMRs because of the stronger Coriolis
force experienced by the horizontally expanding crest of the rising
tubes.  
The animation of the surface field distribution 
is given in a supplementary 
animation (see the online appendix, Fig.~\ref{fig:bsurf}, middle panels). 
Figure~\ref{fig:9dall}d shows that the cycle
signal introduced by the underlying dynamo is largely washed out in the
time evolution of the total magnetic flux at the surface. This is caused
by 1) a larger overlap between consecutive cycles owing to a
stronger $\alpha$-effect, and 2) strong polar magnetic fields in
antiphase with the cycle of emerging flux.

\subsection{The case of $P_{\rm rot}=2~{\rm d}$}
\label{sec:2d}

\begin{figure*}
\centering
\begin{minipage}{\linewidth}
\includegraphics[width=0.45\linewidth, bb=28 28 532 388]{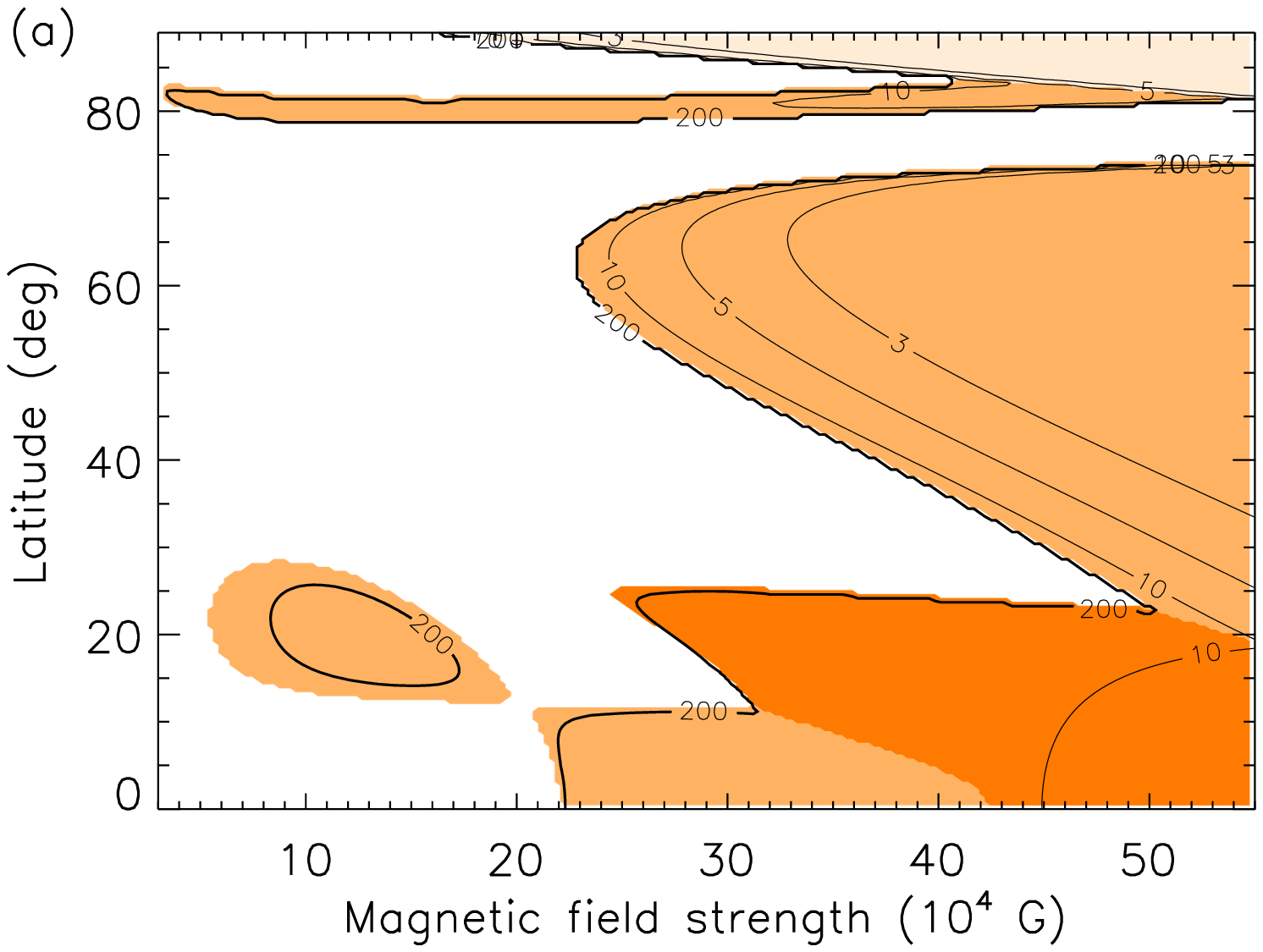}
\includegraphics[width=.5\linewidth]{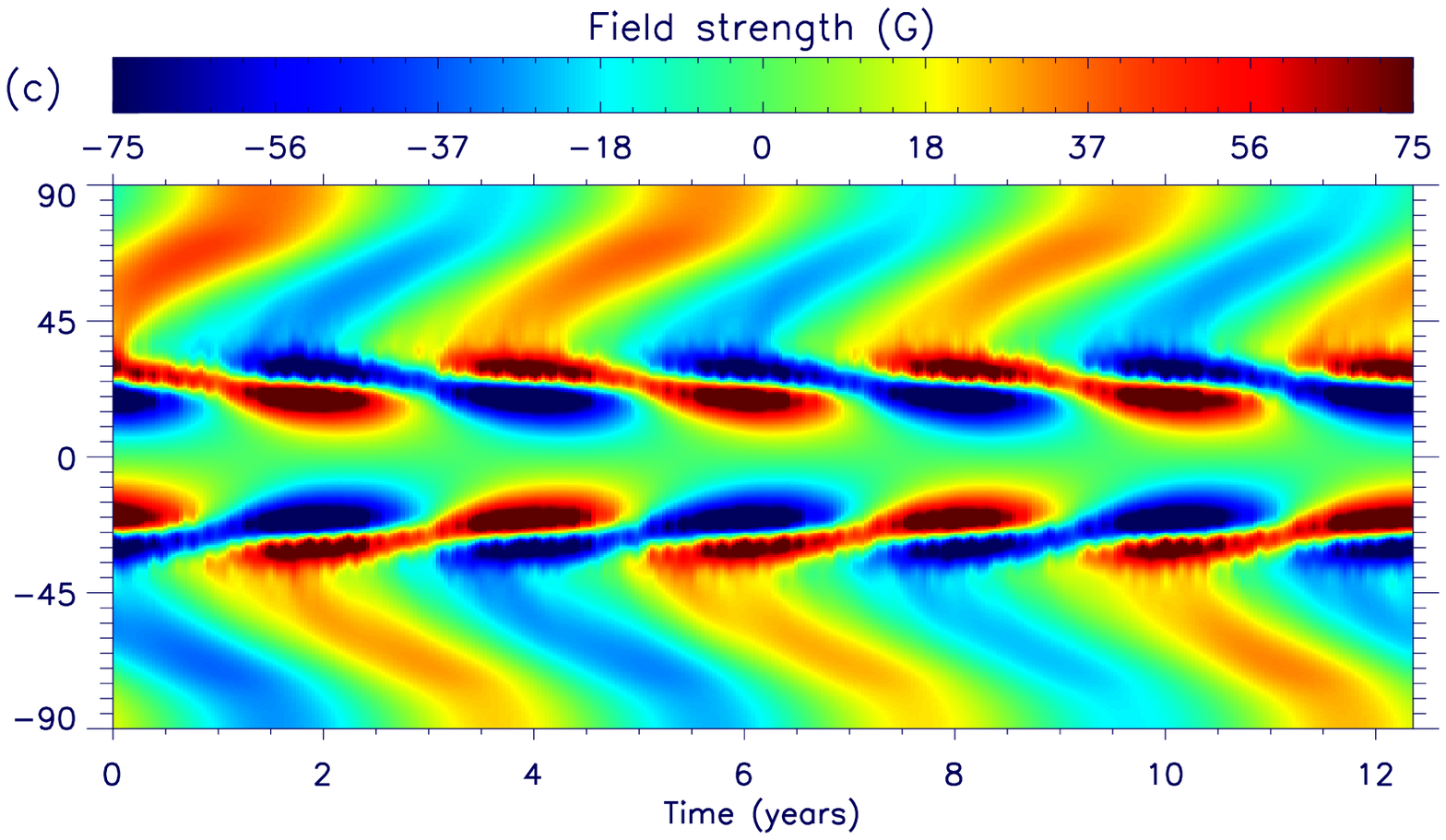} \\
\includegraphics[width=.45\linewidth]{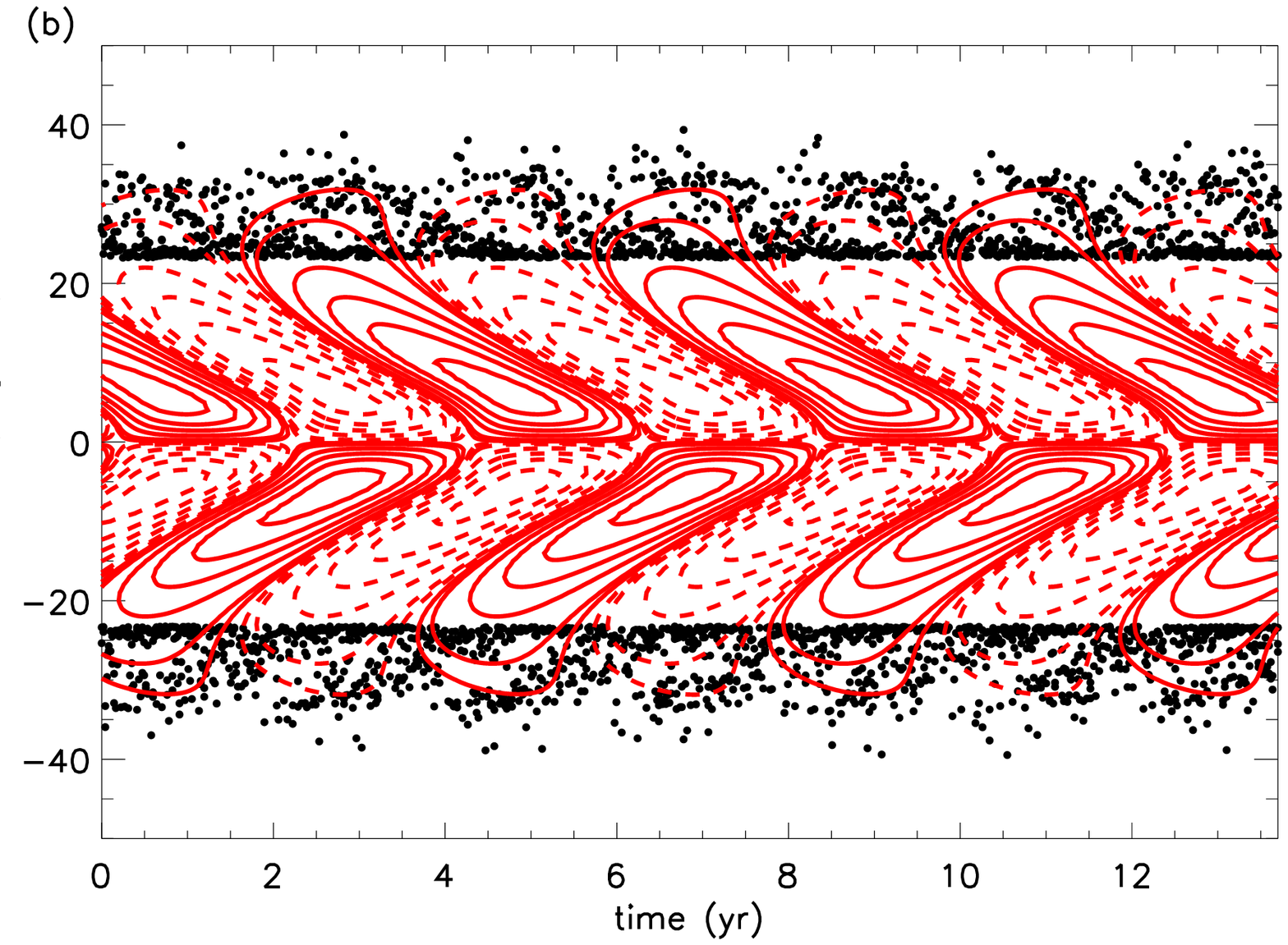}
\includegraphics[width=.5\linewidth]{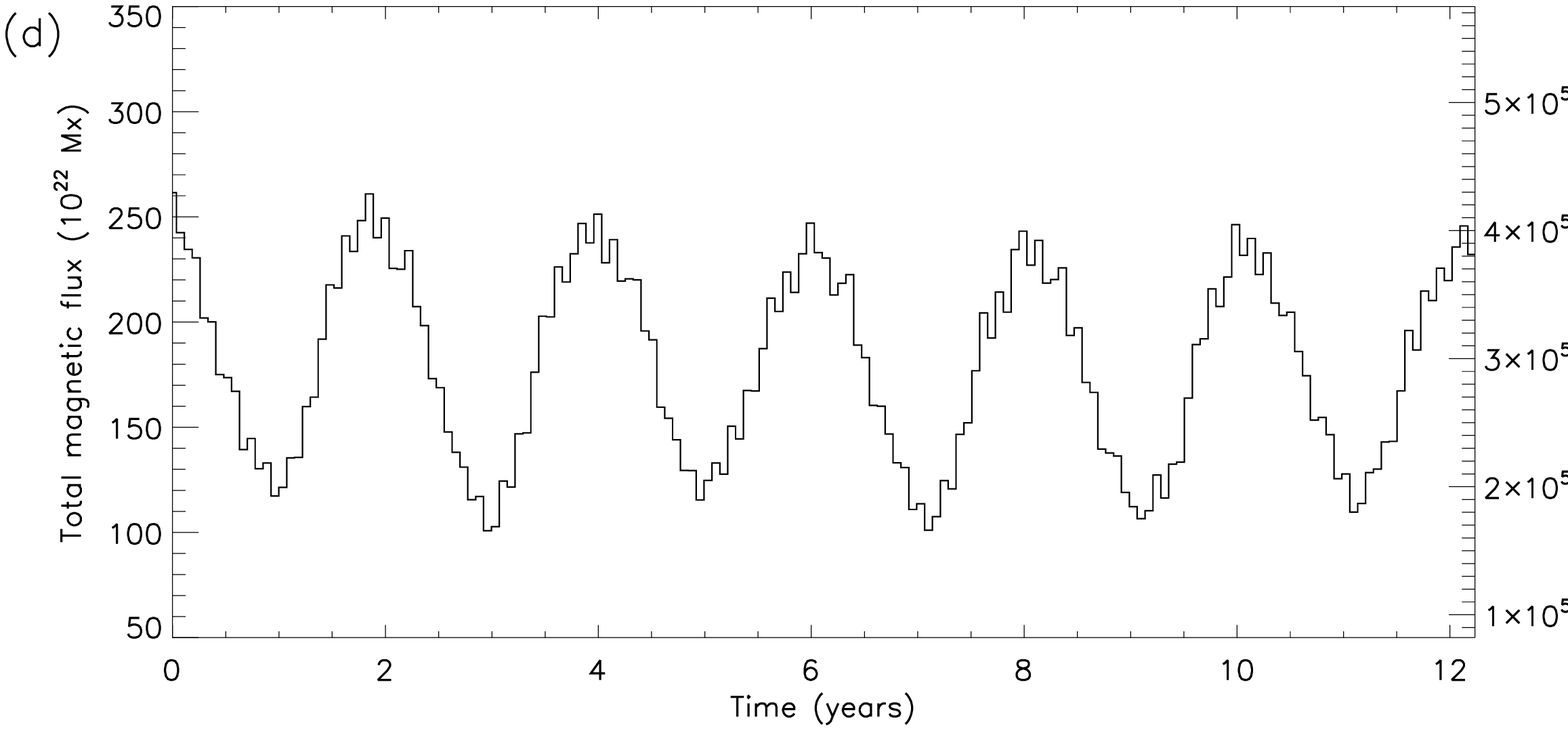}
\end{minipage}
\caption{Same as Fig.~\ref{fig:sunall}, but for a Sun-like star with 
$P_{\rm rot}=2$~d. 
In panel (b), the dots are shown at one-month intervals for clarity.}
\label{fig:2dall}
\end{figure*}
For a star with a rotation rate of $\Omega_\star=13\Omega_\odot$,
corresponding to a period of two~days, the stabilising effect of rotation
on the flux tubes is even stronger than in the previous case. The
stability diagram given in Fig.~\ref{fig:2dall}a shows that the critical
field strength increased to a few times $10^5$~G.  As before, we exclude
the instability islands in determining the field strengths of rising flux
tubes.

For such a large rotation rate, the flux tubes rise almost parallel to
the rotation axis and emerge at a significantly higher latitude than
their starting location \citep{ss92}. The initial
latitudes are in the range of roughly $0^\circ$-$25^\circ$, while the
emergence latitudes are about $23^\circ$-$40^\circ$, so that there is no
flux emergence in an `avoided zone' up to $\pm 23^\circ$ latitude from
the equator.  Figure~\ref{fig:2dall}b shows the dynamo-generated toroidal
magnetic field at the bottom of the convection zone together with the
distribution of flux tubes emerging at the surface. It is evident that
the poleward deflection of flux tubes causes a significant deviation
between the dynamo waves in the interior and the pattern of surface
emergence.

For such a rapidly rotating star, the tilt angles of emerging bipolar
regions are much larger (around $35^\circ$) than in the solar case.  The
relatively narrow latitudinal extension of flux emergence and the large
tilt angles lead to a pair of almost non-migrating latitudinal belts of
opposite magnetic polarity in each hemisphere (see Fig.~\ref{fig:2dall}c
and the supplementary animation in the online appendix, 
Fig.~\ref{fig:bsurf}, bottom panels).
The net poleward transport of flux from the higher-latitude bands produces 
polar fields reaching field strengths of up to 35~G. 
Compared to the $P_{\rm rot}=9$~d case, there is smaller overlap 
between the emergence patterns of consecutive cycles. Consequently, 
the corresponding surface-integrated unsigned magnetic
flux (Fig.~\ref{fig:2dall}d) shows a clearer cycle 
(mainly determined by the activity bands), with a period of 
about two~years, an amplitude of about $1.5\times10^{24}$~Mx, and 
a minimum-to-maximum flux variation by a factor of about two.

\section{Rapidly rotating K stars}
\label{sec:kstars}

\subsection{K0 dwarf}
\label{ssec:k0}

We consider a K0V star at the zero-age main sequence, with a rotation
period of 2 days. Such a star has a smaller radius and
a relatively deeper convection zone than the Sun. We use a stellar model
with $R_{\star}=0.735R_{\sun}$ and the bottom of the convection zone
located at $0.677R_{\star}$ \citep{voho01}.

\begin{figure*}
\centering
\includegraphics[width=0.45\linewidth, bb=28 28 532 388]{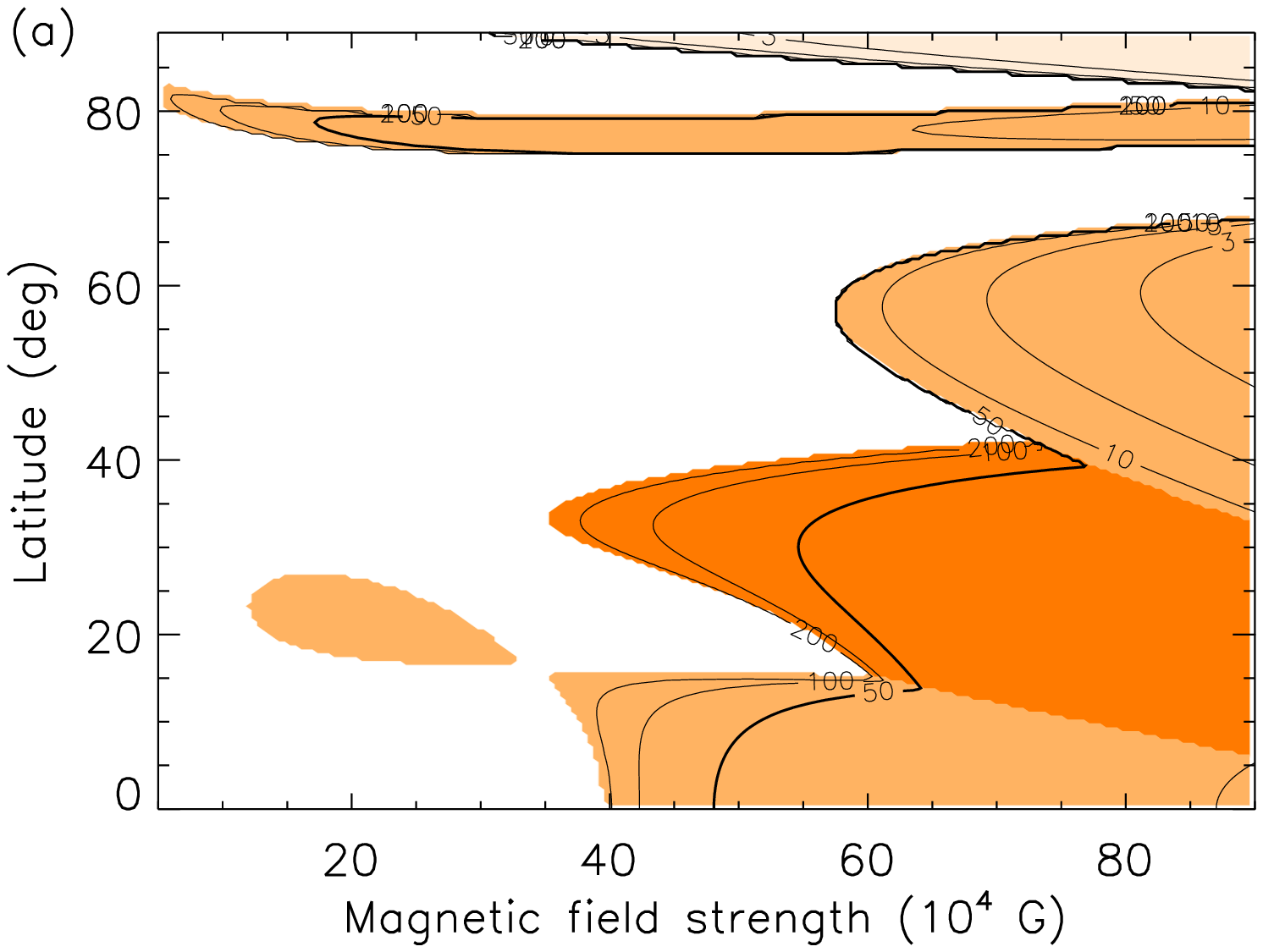}
\includegraphics[width=.5\linewidth]{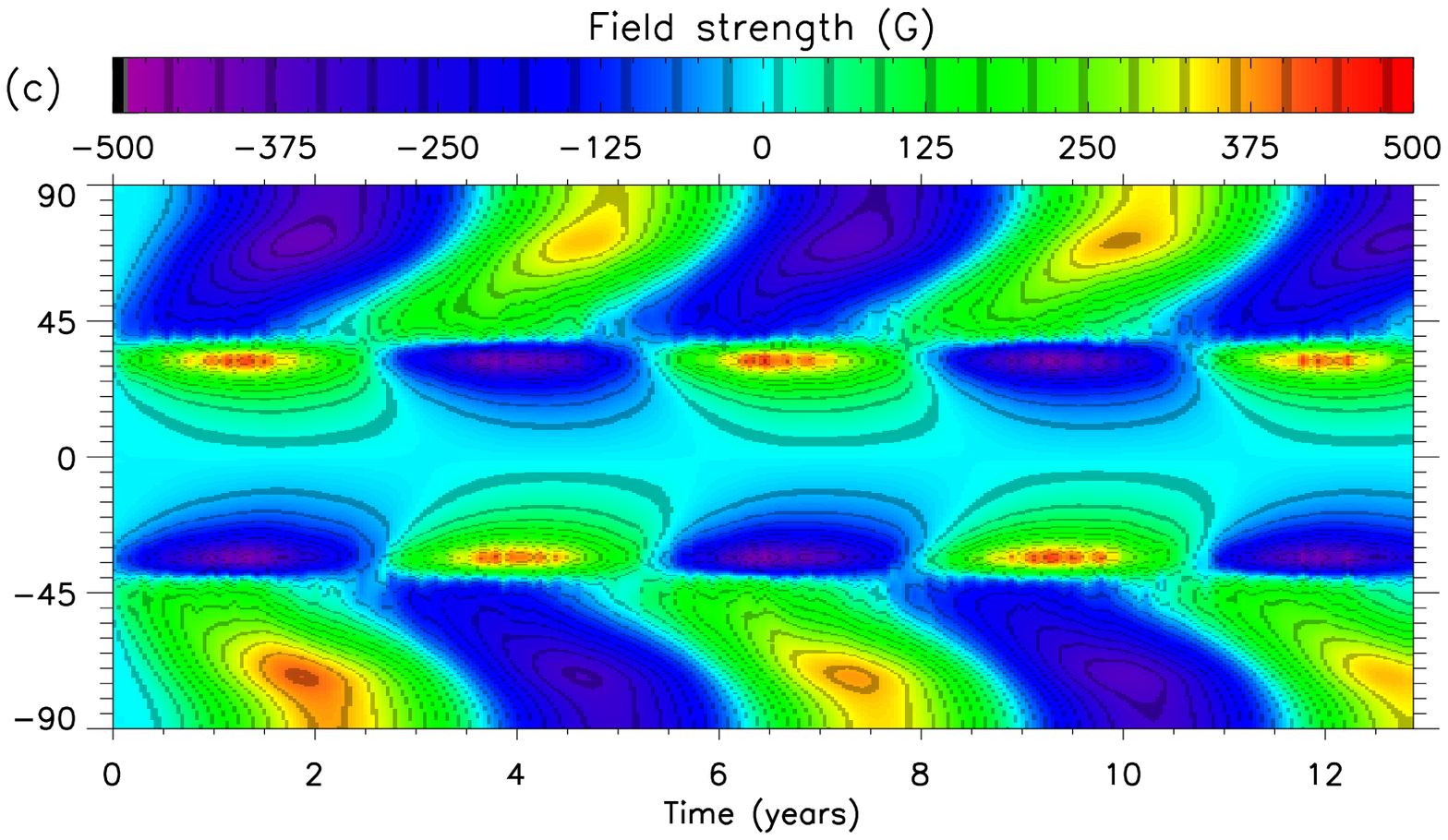} \\
\includegraphics[width=.45\linewidth]{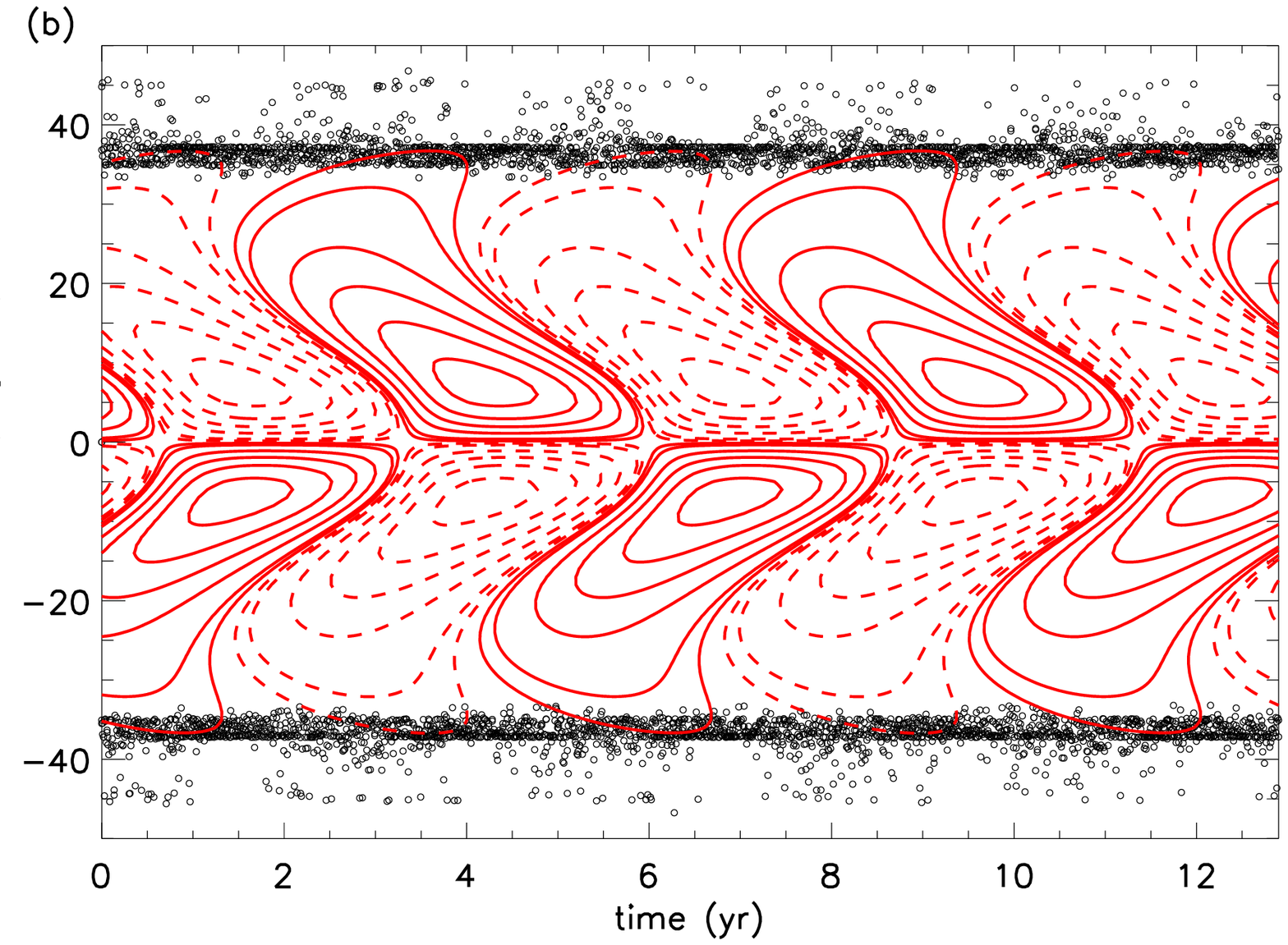}
\includegraphics[width=.5\linewidth]{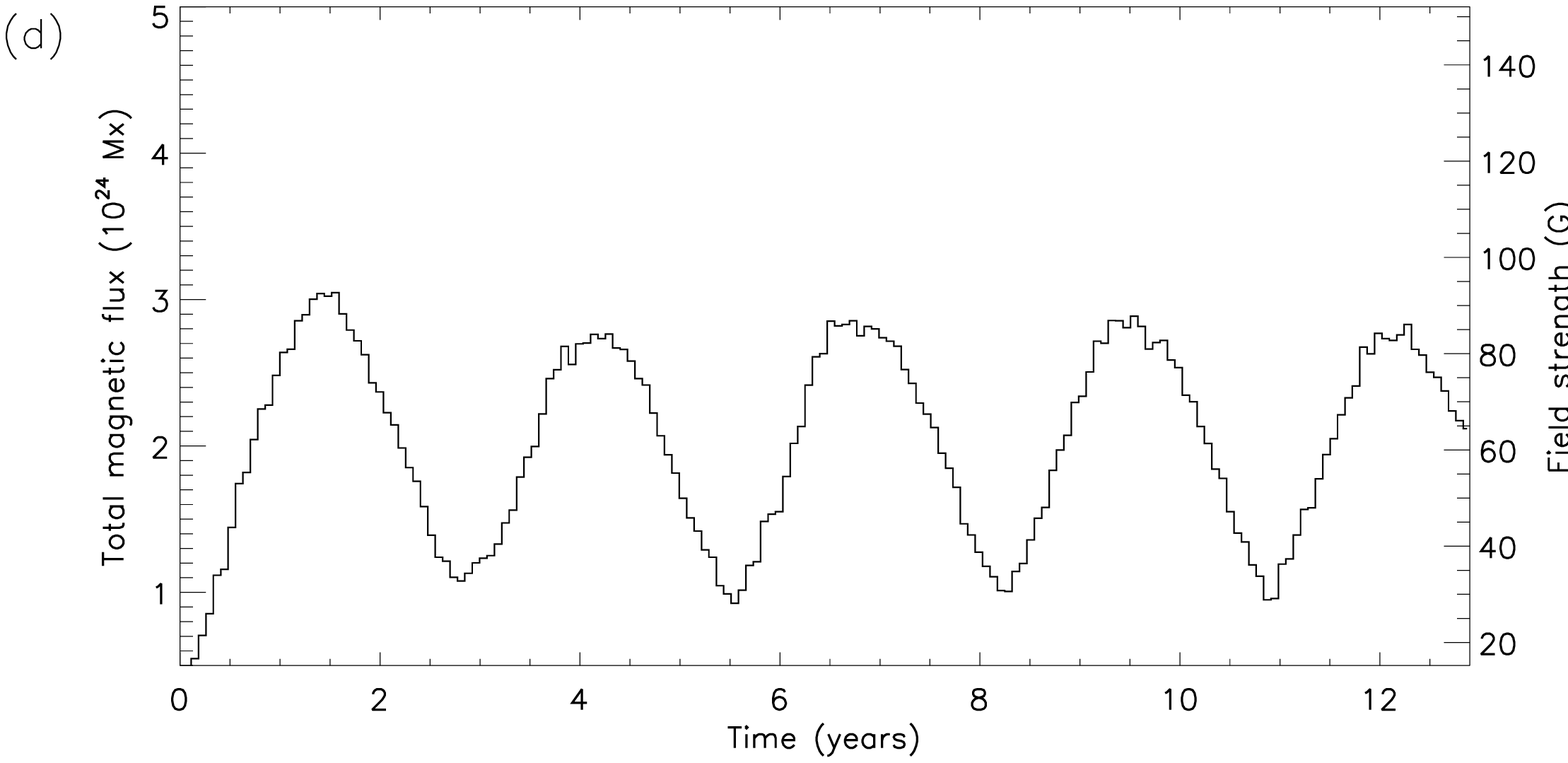}
\caption{Same as Fig.~\ref{fig:sunall}, but for a K0V star with 
$P_{\rm rot}=2$~d.
}
\label{fig:k0vall}
\end{figure*}

The Reynolds numbers for differential rotation, $R_\Omega$, and the
$\alpha$-effect, $R_\alpha$, have to be adapted to the new value of the
length scale, namely the radial distance of the convection zone lower
boundary from the center.  Scaling with $\Omega_\star/\Omega_{\sun}$, we
obtain $\alpha_0=152$~cm~s$^{-1}$. The resulting dynamo number $R_\alpha
R_\Omega$ is about a factor of four smaller than in the case of the Sun-like
star with $P_{\rm rot}=2$~d (Sect.~\ref{sec:2d}), which leads to a
somewhat longer magnetic cycle period of about 6.5 years.  The number of
emerging flux tubes per cycle is set to the same value as in the
Sun-like star with the same rotation period. The emergence latitudes and
tilt angles of the BMRs are determined with the same procedure as
in the previous sections, considering the stability properties of flux
tubes in the mid-overshoot region of the K0V star
(Fig.~\ref{fig:k0vall}a).

The comparison of the dynamo waves of the deep-seated toroidal field
with the emergence pattern at the surface is shown in
Fig.~\ref{fig:k0vall}b.  The difference between the emergence latitudes
and the initial latitudes of flux tubes is larger by about $10^\circ$
than for the solar-type star with the same rotation rate. This is due to
a simple geometric effect: the K0V star has a deeper convection zone (in
terms of fractional radius), so that a flux loop rising parallel to the
rotation axis emerges at a higher latitude than in the case of a
shallower convection zone.

The time-latitude diagram of the longitudinally averaged magnetic field
and the time variation of the integrated unsigned surface flux are shown
in Fig.~\ref{fig:k0vall}c. Owing to the rapid rotation, the emerging BMRs
have large tilt angles between $40^\circ$ and $55^\circ$. This leads to
pronounced branches of follower polarity drifting poleward via 
diffusion and meridional circulation. 
The cycle signal is similar in strength to the corresponding 
Sun-like case (Sect.~\ref{sec:2d}), 
again due to the small cycle overlap of the emergence patterns 
and the large tilt angles leading to the formation of activity bands 
(Fig.~\ref{fig:k0vall}d). 

\begin{figure}
\centering
\includegraphics[width=.33\columnwidth]{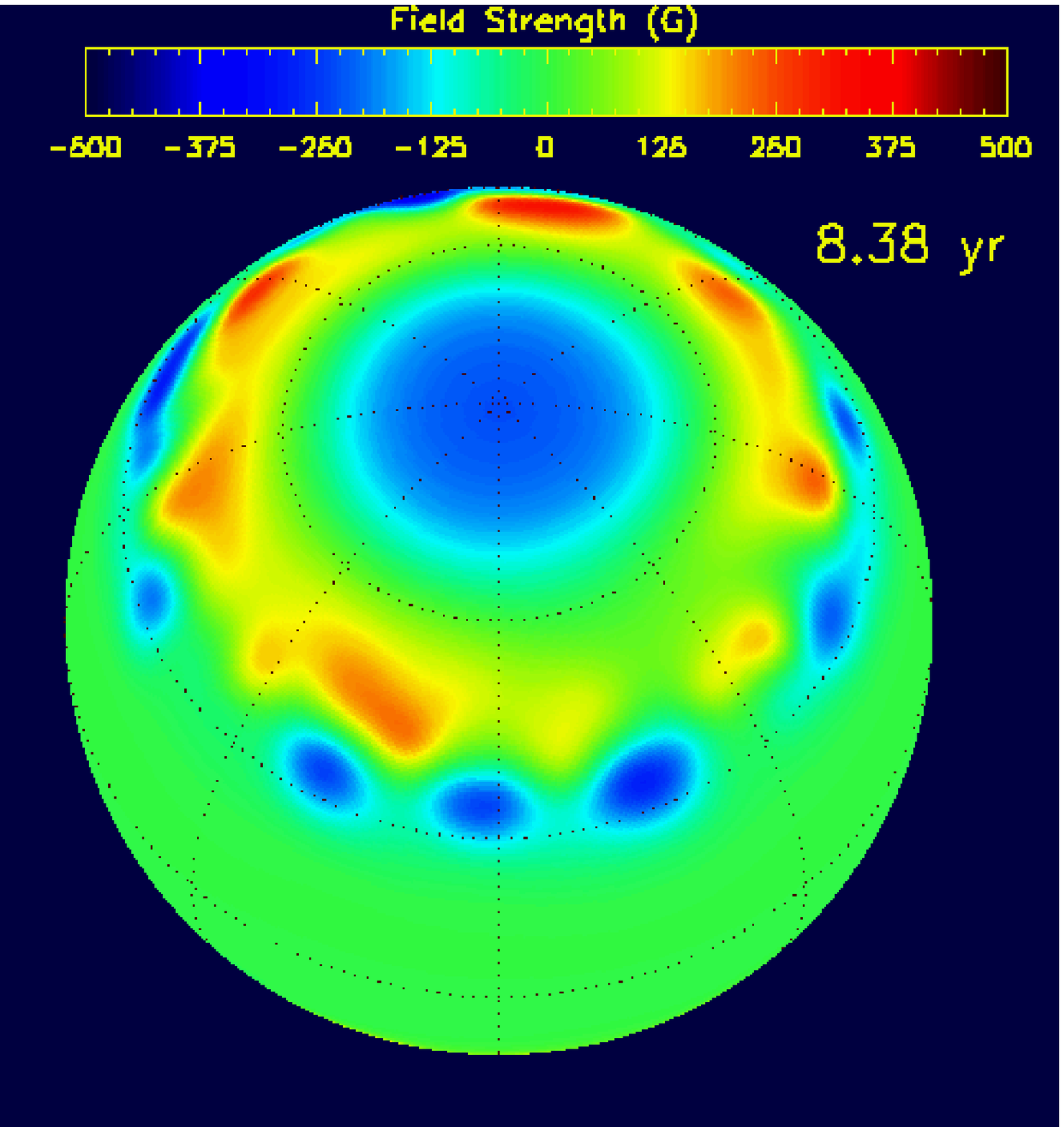}\includegraphics[width=.33\columnwidth]{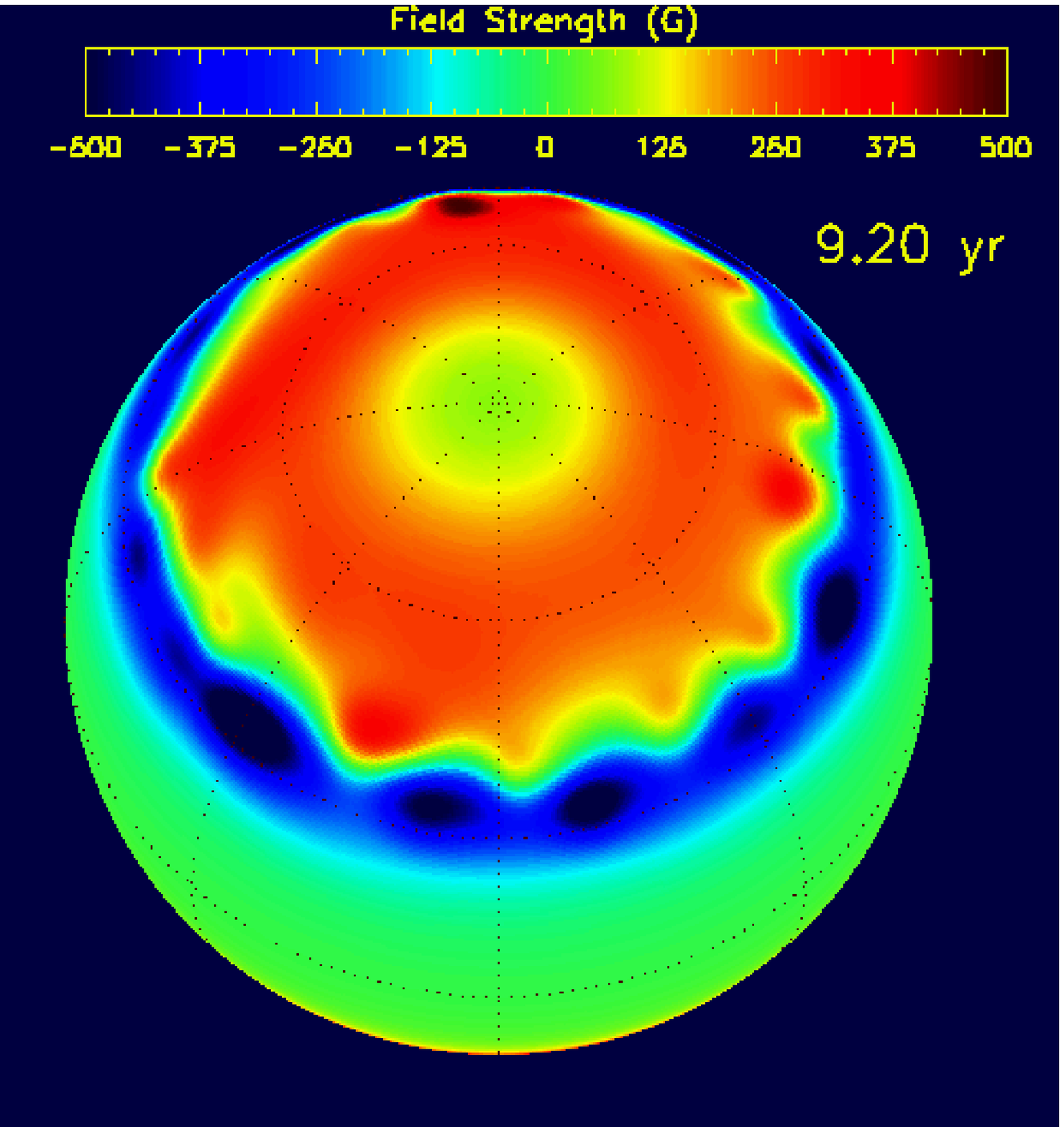}\includegraphics[width=.33\columnwidth]{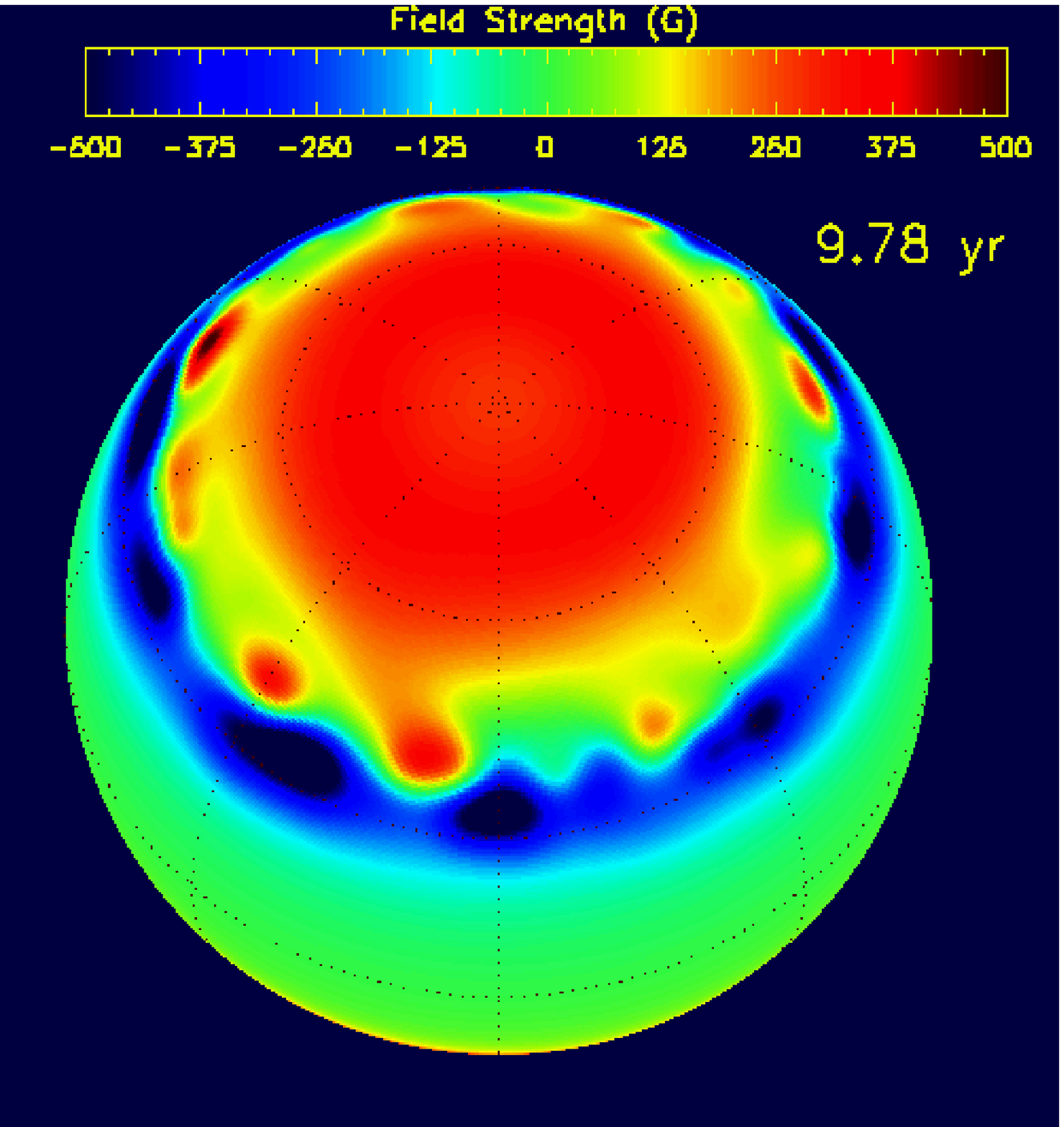}
\caption{Snapshots of the magnetic field distribution at the surface of 
the K0V star with $P_{\rm rot}=2$~d. The colour table shows the field
strength with a saturation level at $\pm 500$~G. 
\emph{Left:} the activity minimum phase. \emph{Middle:} 
the activity maximum phase. \emph{Right:} the beginning of the 
decay phase following the activity maximum, when the polar field reaches 
its maximum strength. The
inclination of the rotation axis with respect to the line of sight is
$30^\circ$.}
\label{fig:K0Vsnap}
\end{figure}

Three snapshots of the flux distribution at the stellar surface are
shown in Fig.~\ref{fig:K0Vsnap}. 
Near an activity minimum (left panel),
mid-latitude BMRs of the newly started cycle lead to an uneven 
longitudinal distribution of magnetic flux, owing to the relatively low 
emergence frequency. In the maximum phase (middle panel), 
highly tilted BMRs lead to the formation of a pair of latitudinal belts of
opposite polarity in each hemisphere. The higher-latitude belt provides
the source for the respective polar field of the same polarity. After seven
months (right panel), flux emergence has decreased in parallel to the
evolution of the deep-seated toroidal field generated by the dynamo, 
while the polar field has reached its maximum strength. 
As a result, the
surface-integrated unsigned magnetic flux exhibits a cyclic variation
with a period of about 2.75~yr (Fig.~\ref{fig:k0vall}d). 

\subsection{K1 subgiant}
\label{ssec:k1sub}
As a further example, we consider a K1 subgiant rotating with a period 
of $2.8$~days. A star of this type is
the magnetically active component of HR 1099, the most thoroughly
studied RS CVn-type binary system. In the structure model adopted here
\citep{voho01}, the stellar mass is $1.5 M_{\sun}$, the age is about
$2.4$~Gyr, the stellar radius is $R_{\star}=3.44 R_{\sun}$, and the
bottom of the convection zone is at $0.35 R_{\star}$. These parameters
still allow flux tubes to rise up to the surface. If the bottom of the
convection zone were located deeper than about $0.30 R_{\star}$, buoyant
flux tubes entering the convection zone would be trapped within the star
as a consequence of the magnetic curvature force \citep{voho01}.  

\begin{figure*}
\includegraphics[width=0.45\linewidth, bb=28 28 532 388]{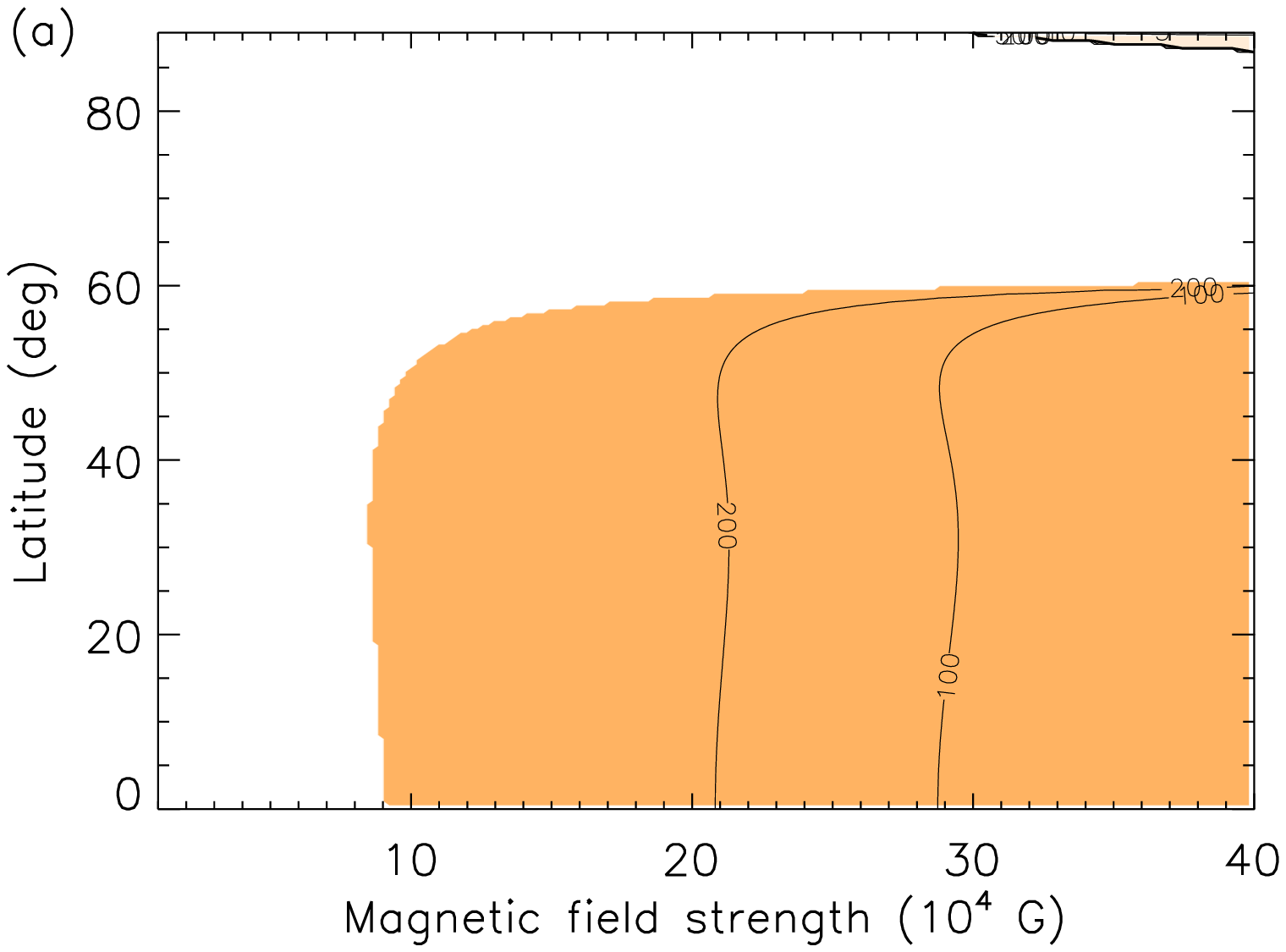}
\includegraphics[width=.5\linewidth]{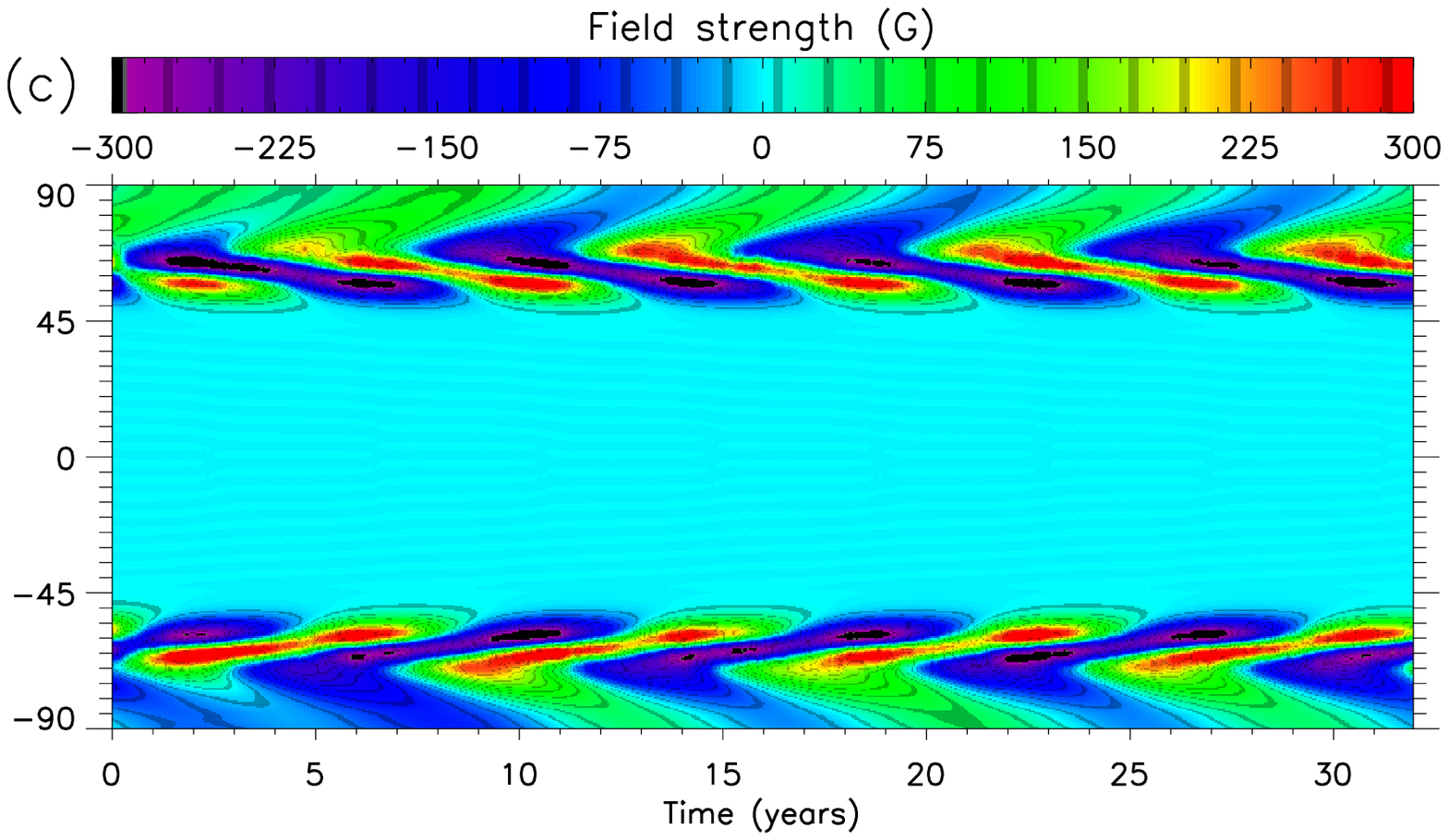} \\
\includegraphics[width=.45\linewidth]{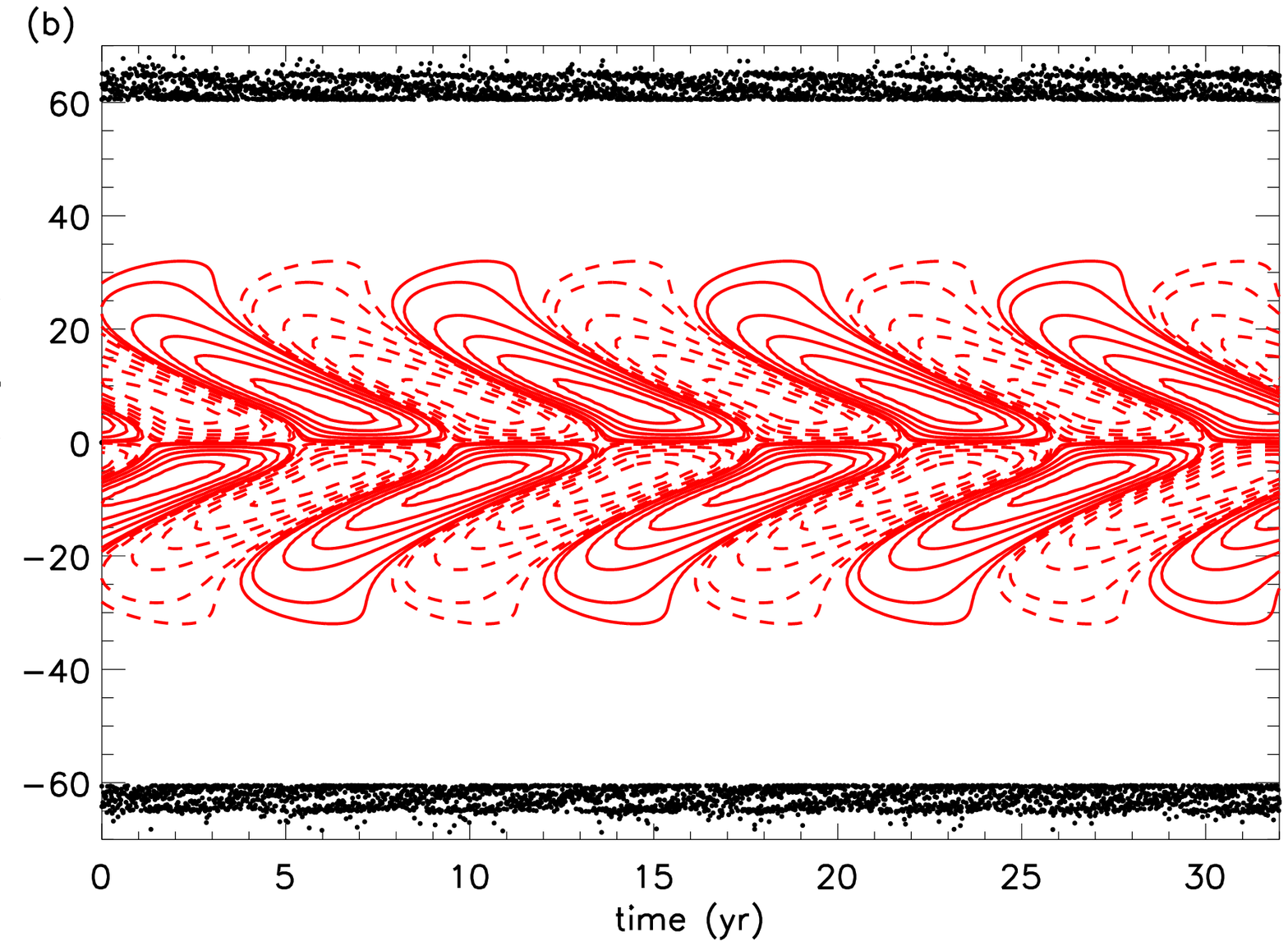}
\includegraphics[width=.5\linewidth]{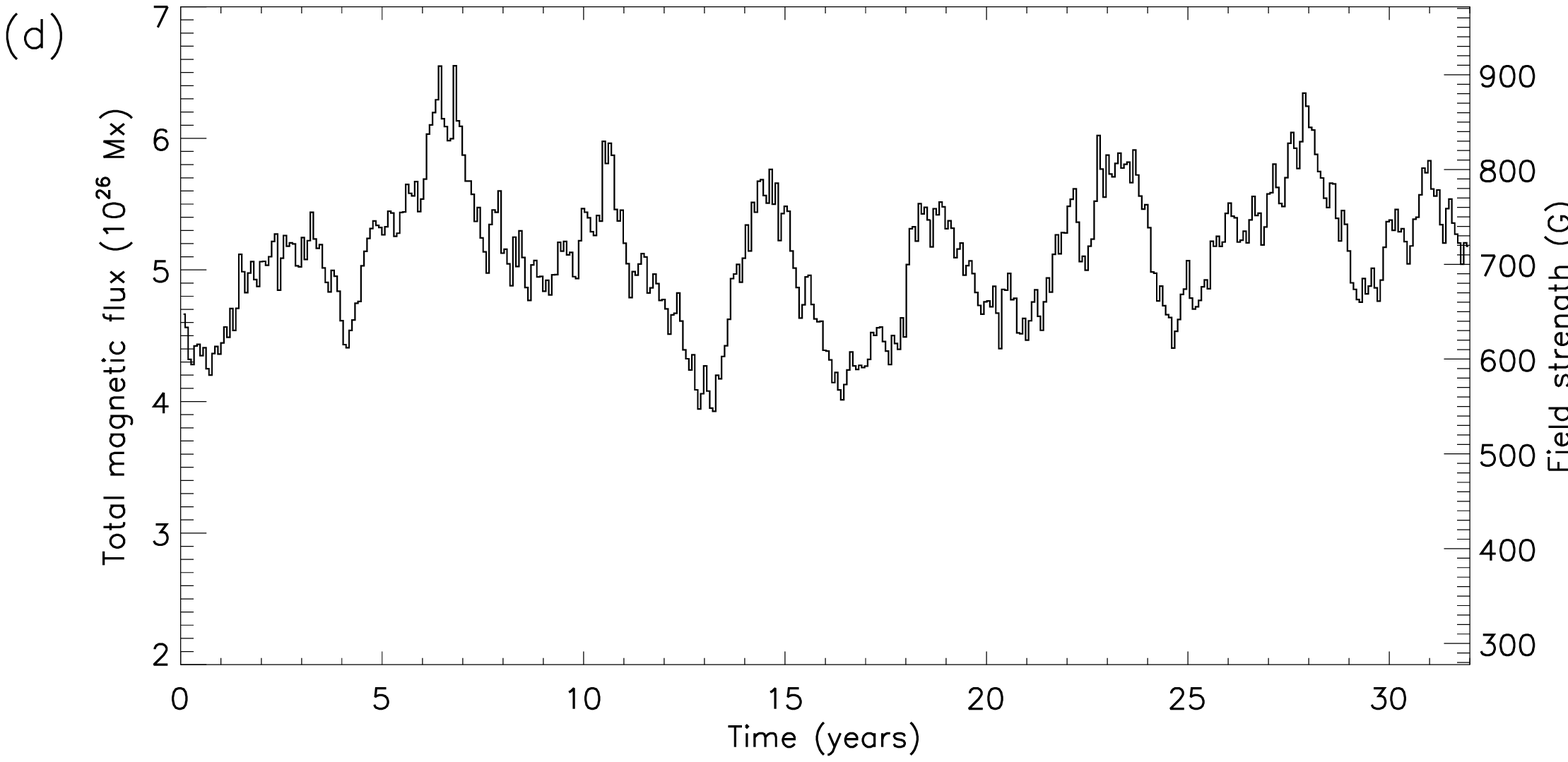} 
\caption{Same as Fig.~\ref{fig:sunall}, but for a K1 subgiant star with 
$P_{\rm rot}=2.8$~d and rigid rotation. }
\label{fig:k1all}
\end{figure*}

We assumed rigid rotation in the calculations 
regarding the instability and the rise of magnetic flux tubes, 
and the surface flux transport. 
This assumption is supported by the very weak surface 
shear observed in HR 1099 \citep[][]{don03b,petit04}. 
Furthermore, the effects on flux tube dynamics 
of differentially rotating convection zones in post-main sequence stars 
is yet to be explored, as an extension of the work by \citet{voho01}. 
The linear stability diagram for 
flux tubes in the middle of the overshoot region (about $0.33 R_{\star}$) 
is shown in Fig.~\ref{fig:k1all}a. 

For the dynamo model, $R_\Omega$ has been rescaled using the radius 
$r_0=1.1 R_\odot$ 
of the star's radiative core and the surface latitudinal shear of about 
$0.87^\circ {\rm d}^{-1}$, following the measured value with the lowest 
variance, as given by \citet{petit04}. The resulting dynamo 
number leads to an activity cycle with a period of about 4 yr, 
compared to the observed cycle periods of 5.3 yr and 16 yr \citep{berdyugina07}. 


Figure~\ref{fig:k1all}b shows that the dynamo waves and the emergence
pattern have no spatial correlation. As a result of the deep convection
zone and the geometric effect described in Sect.~\ref{ssec:k0}, the
average latitude of emergence is shifted to even higher latitudes than
in the case of main-sequence stars.  The emergence belt centred around
$\lambda\simeq 62^\circ$ has a latitudinal width of only about 7 degrees.
This is caused by the projection of a 
low-latitude angular range at the bottom of the convection zone, onto the
surface and along the direction parallel to the rotation axis, producing 
a much narrower range at high-latitudes. 

Emerging BMRs have tilt angles between $5^\circ$ and $8^\circ$, which are
significantly smaller than those for the rapidly rotating main-sequence stars
considered in the previous sections. This can be explained in the present case 
by the rising flux tubes spending a much longer time
(about five years) in the convection zone, so that the magnetic tension
force can more efficiently act against the twisting effect of the Coriolis
force on the horizontally expanding loop.

The time-latitude diagram of the longitudinally averaged magnetic field
strength (Fig.~\ref{fig:k1all}c) is similar to that of
the rapidly rotating Sun-like star, apart from a shift of the
patterns to higher latitudes. 
The total unsigned surface flux variation (Fig.~\ref{fig:k1all}d) shows
a rather weak cyclic signal upon which irregular variations are
superimposed. The narrow latitude range of flux emergence, as well as
the weak transport effects by surface flows and diffusion in this region,
amplify the effect of the random component of the surface flux injection
process. 
This results in rather large fluctuations of the total unsigned surface flux.

\begin{figure}
\centering
\includegraphics[width=.45\linewidth]{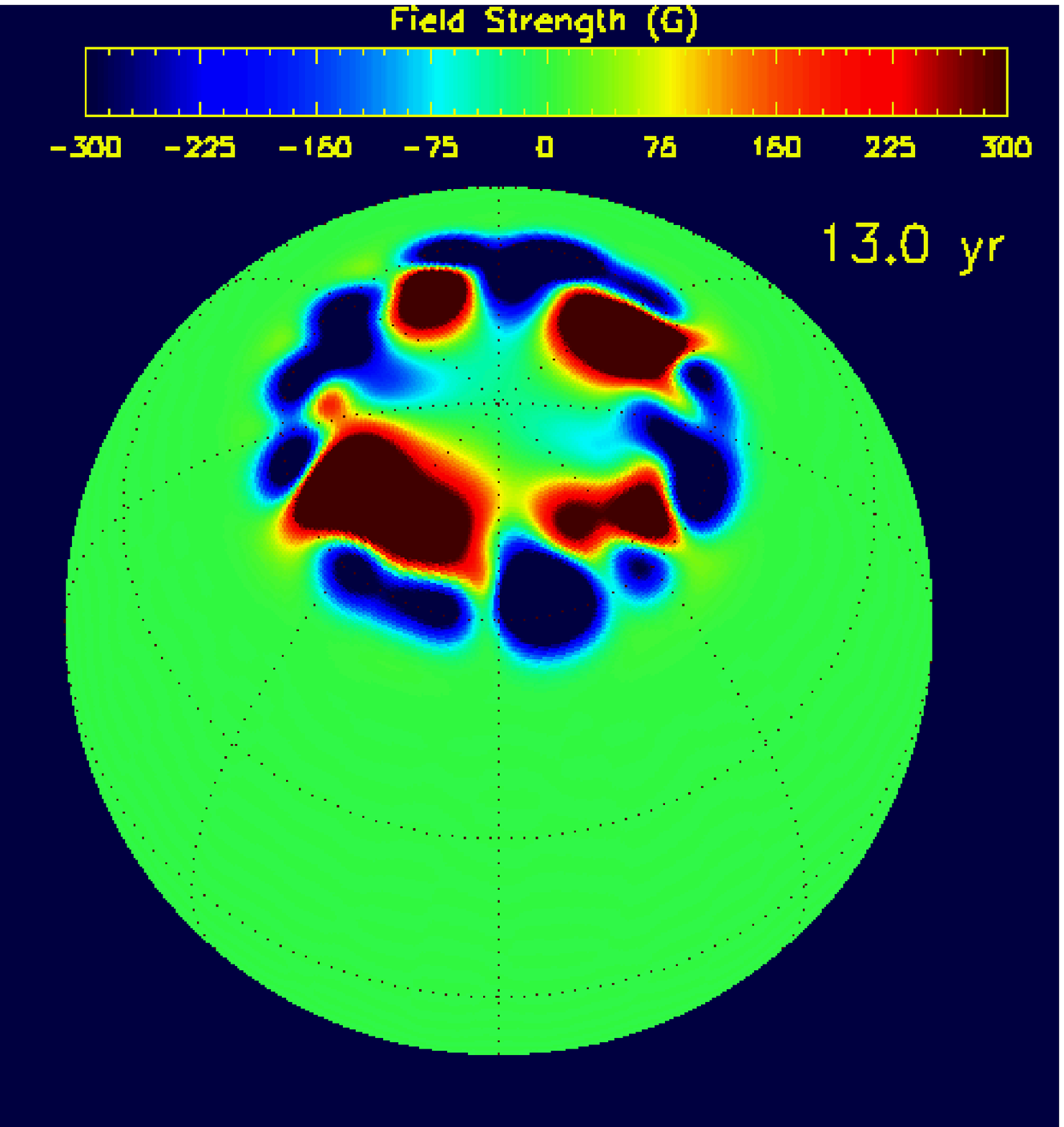}\includegraphics[width=.45\linewidth]{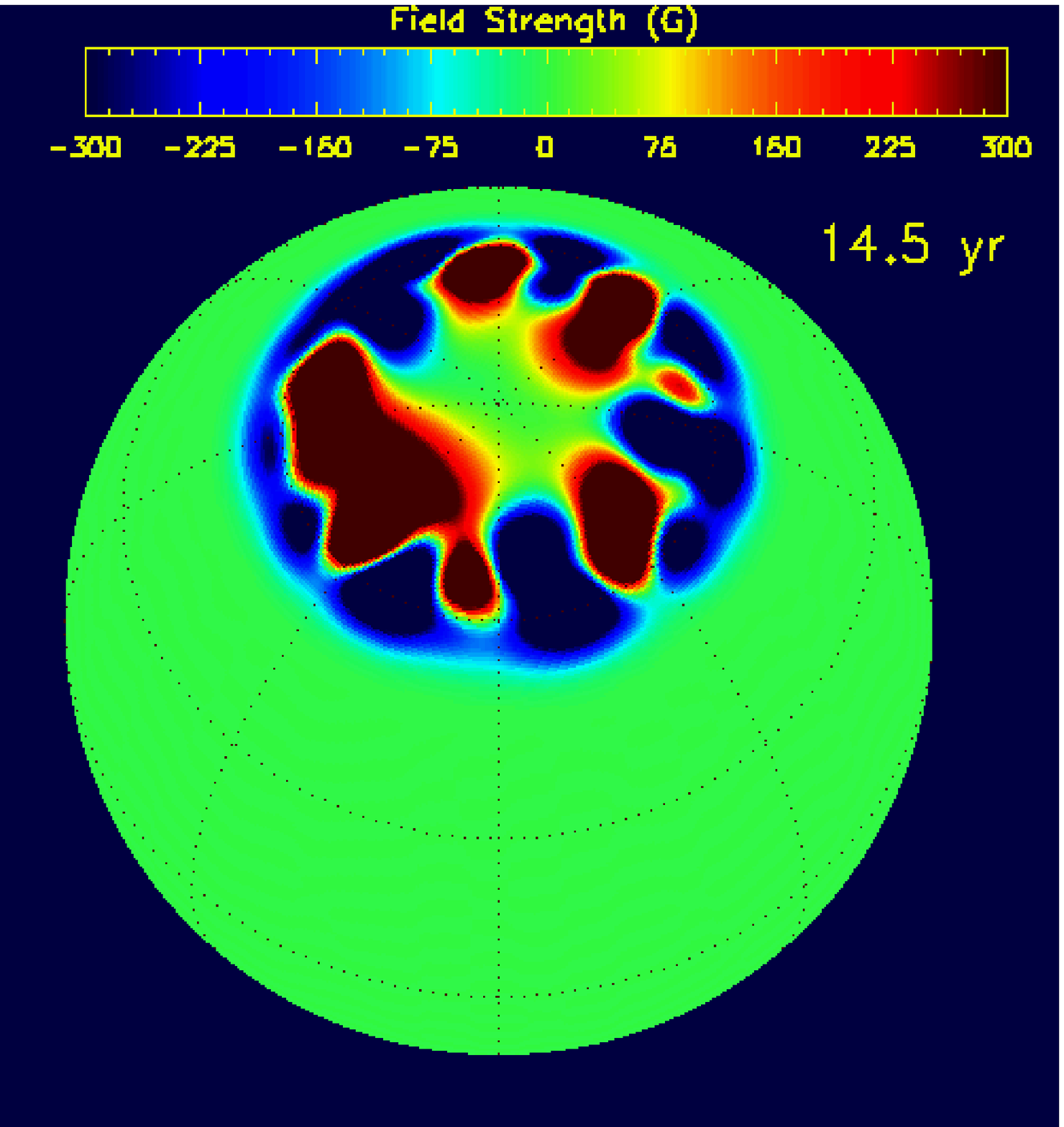}
\caption{Field strength distribution at an activity minimum (left panel), 
and at a maximum (right panel), for the K1 subgiant star.  The
inclination of the rotation axis with respect to the line of sight is
$30^\circ$. The colour table shows the field
strength with a saturation level at $\pm 300$~G.} 
\label{fig:minmax}
\end{figure}

Figure~\ref{fig:k1all}c also indicates a rather weak field in the polar
caps above $\sim75^\circ$ latitude, although polarity reversals still
occur. This weakness is due to the relatively small tilt angles of the
BMRs and the long diffusion timescale (proportional to $R_\star^2$): the
meridional flow vanishes beyond $\pm75^\circ$, so that further poleward
flux transport occurs only by means of the relatively inefficient diffusion
process. Inefficient diffusion and small tilt angles lead to an
intermingling of polarities in the narrow flux emergence belts, as shown
in Fig.~\ref{fig:minmax}. Therefore, the resulting poloidal field
produced by the diffusing BMRs is weak and far from a simple dipole
configuration. 

\section{Discussion}
\label{sec:disc}

The combined model of magnetic field generation and transport provides
two major improvements compared to previous models \citep[e.g.,][]{bsss04,
voho06}: 

\begin{enumerate}
\item The cycle properties are determined by an underlying
dynamo model, whereas in previous studies either the butterfly diagrams
were prescribed arbitrarily, or observed emergence patterns were
used. We thus establish the nontrivial link between the deep-seated
field generation and surface flux transport.
\item The tilt angles and emergence latitudes of bipolar magnetic
regions at the surface are not prescribed, but consistently determined
from simulations of rising flux tubes.  This may strongly affect the
surface distribution of magnetic flux, and particularly the strength of
the polar magnetic fields.
\end{enumerate}

In this first exploratory study, our main motivation was to
evaluate the effects of combining models for the dynamo, flux tube
stability and dynamics, and surface transport, rather than to constrain
dynamo models for cool stars. We therefore adopted a rather simple
thin-layer dynamo model with differential rotation in radius, which does
not include the probably important effects of latitudinal differential
rotation and meridional circulation. We note also that \citet{Spruit:2010}
argues that the radial angular velocity gradient in the solar tachocline
is not capable of building up a sizeable toroidal field. Our
model can certainly be utilised to test more elaborate dynamo models, such as 
Babcock-Leighton-type flux transport dynamos \citep{char05}.


For solar-type and K0 main-sequence stars, we find that the polar
magnetic fields for rotation periods of 10~d and 2~d are stronger than
in the solar case, indicating the importance of surface transport
effects \citep[][]{st01,isik07b}: the emergence and meridional
transport of BMRs at mid-latitudes with tilt angles as high as
$35^\circ$ lead to strong (in comparison to the Sun) unipolar flux 
regions around the poles.

For the solar-type model with $P_{\rm rot}=10$~d, the surface flux 
transport blurs the periodic
signal of the oscillatory dynamo in the surface-integrated unsigned
flux.  This suggests that for rapid rotators, a cyclic dynamo may
actually underly an apparently irregular variation in activity
indicators.  This could possibly be the case for HD~190771, a Sun-like
star with $P_{\rm rot}=8.8$~d observed by \citet{petit09}. Using the
Doppler-imaging technique, they report that the total magnetic energy at
the surface does not change over the observed period, although global
polarity reversals in the radial and azimuthal field components
occur. Although considering only a radial field at the surface, our
models show that a roughly constant total magnetic flux at the surface
may arise from the combined effects of dynamo cycle overlap, large tilt
angles, high flux emergence frequency, and meridional transport.

For all models with $P_{\rm rot}=2$~d, we found that the dynamo wave
pattern and the emerging surface flux are completely different from each 
other, owing to
the strong poleward deflection of rising flux loops by the Coriolis
force.  Consequently, for rapid rotators the observed activity patterns
on the surface may not necessarily be taken to represent the
spatio-temporal distribution of magnetic field in the deep interior.

The results presented here are certainly rather preliminary, owing to
the very simple dynamo model used and the insufficient information
we have at present about various important parameters:

\begin{itemize}

\item The dependence on rotation rate and stellar structure of
$\alpha$-effect, internal rotation profile, and the meridional flow
velocity are largely unknown.

\item Various complications regarding the formation of flux tubes, the
growth of the buoyancy instability, and the rise through the convection
zone are omitted, because the connection between dynamo-generated field
and flux tube dynamics is not well understood \citep[e.g.,][]{afmsch03}.

\item Reliable models for the structure of convective overshoot regions
  do not exist. 


\item It is unclear whether the solar value of
$\eta_h=600$~km$^2$~s$^{-1}$ for the turbulent surface diffusivity is
valid for stars of different structure, rotation, and fraction of
surface covered by magnetic flux.

\end{itemize}

For the solar-type stars with $P_{\rm rot}=10$~d and 2~d and the K0V
star with $P_{\rm rot}=2$~d, low/mid-latitude flux emergence and strong
polar magnetic fields co-exist.  However, the lack of magnetic regions
for $|\lambda| < 60^\circ$ for the K1 subgiant is in disagreement with
observations for which active regions are detected at lower latitudes 
\citep[][]{vogt99,strassbartus00}. 
This could be due to an underestimation of the required
field strength for flux tube instability and rise: if the
subadiabaticity in the overshoot region of the subgiant were higher
than given by the stellar model used, unstable rising flux tubes would
have a stronger field and thus be more buoyant, leading to less poleward
deflection by the Coriolis force and their emergence at lower latitudes.
Another possible explanation is non-solar meridional flow profiles (e.g., with
large multiple cell structures), which can have a large impact on the
cycle properties, particularly for advection-dominated dynamo models
\citep{jouve09}. Furthermore, these patterns could also provide equatorward
flux transport.

Our assumption that the number of flux tubes emerging per activity
cycle is proportional to the rotation rate is qualitatively consistent
with our scaling of the $\alpha$-effect with rotation rate.  However,
there are energetic and geometrical constraints on the maximum amount
of magnetic flux that can be generated in rapid-rotator convection zones
with the known mechanisms \citep{rempel08}.

\section{Conclusions}
\label{sec:conc}

We have presented a combined model including 1) a thin-layer dynamo with an 
  $\alpha$-effect due to magnetic buoyancy instability and radial shear
  according to helioseismic results, 2) the stability analysis and
  dynamical simulation of flux tubes rising through the convection zone,
  and 3) surface flux transport by differential rotation, meridional
  flow, and turbulent diffusion that reproduce the basic observed
  features of the solar cycle (such as the butterfly diagram of flux
  emergence, the evolution of the large-scale surface flux distribution,
  and the polar field reversals). 

Extending the model to rapidly rotating Sun-like and K-type
  main-sequence stars, we have found that the latitude distribution 
of emerging flux at the surface may differ significantly from that of the
  deep-seated toroidal field generated by the dynamo process. A strong
  overlap of the cycles may hide the cyclic signal in activity indicators
  related to the integrated total surface flux. Low- to mid-latitude
  flux emergence and activity can co-exist with strong polar fields
  (`polar spots') arising from poleward surface flux transport. 

In the case of a K1 subgiant, we have found that the large depth of the convection
  zone and long diffusion time lead to belts of intermingled
  mixed-polarity field at high latitudes and relatively weak polar
  field.

For improved realism of the model, a two-dimensional
  flux-transport dynamo model will be incorporated at our next proposed developed 
stage. Information about the stellar interior and the surface
  differential rotation, meridional flow, and internal structure can be
  included in the model as it becomes available by means of
  asteroseismology.

\bibliographystyle{aa}
\bibliography{dynpap}

\Online

\appendix

\section{Animations}

\begin{figure}
\centering
\includegraphics[width=.5\linewidth]{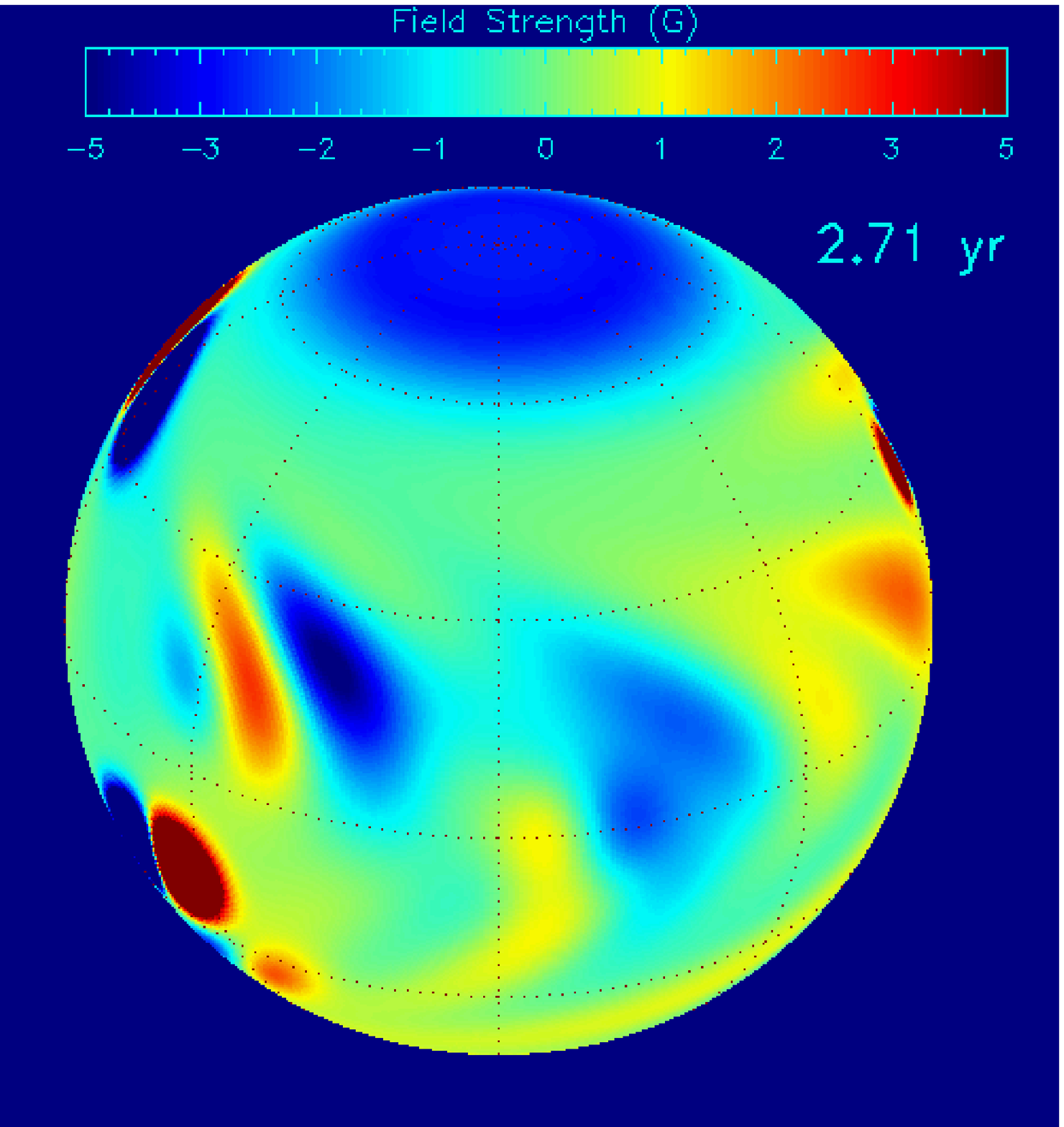}\hskip-1.5mm
\includegraphics[width=.5\linewidth]{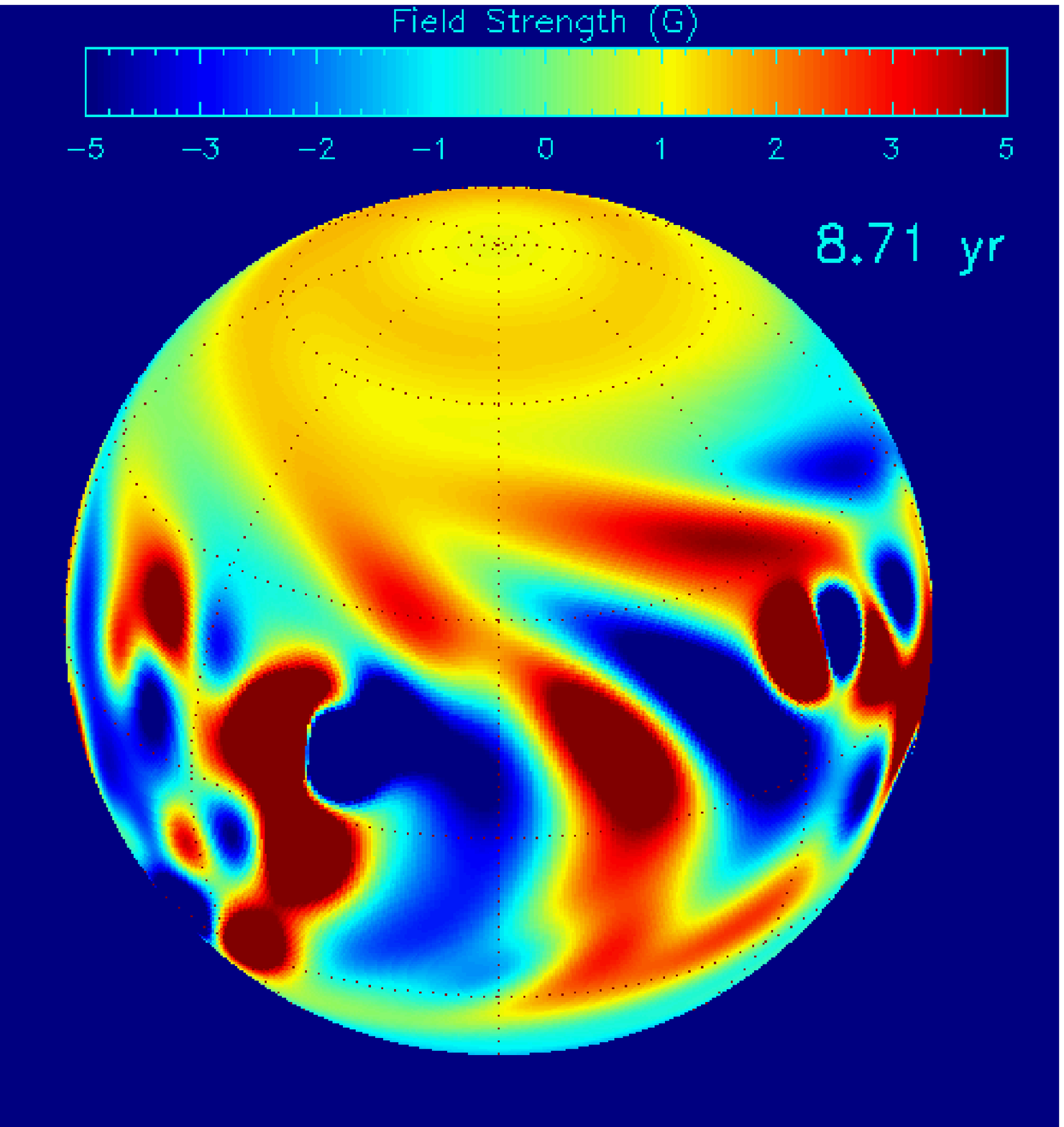}\hskip-1.5mm\\
\vskip2mm
\includegraphics[width=.5\linewidth]{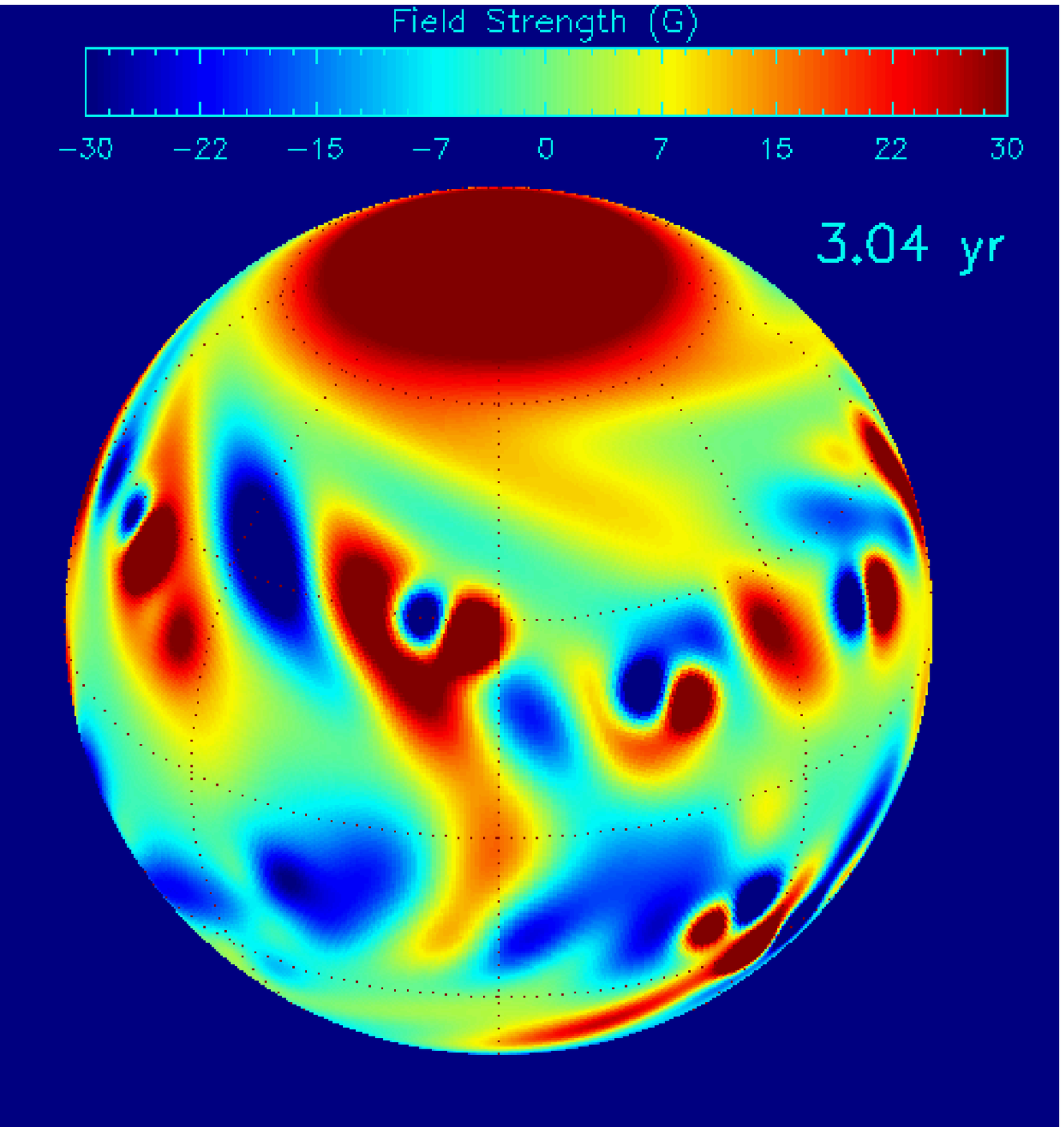}\hskip-1.5mm
\includegraphics[width=.5\linewidth]{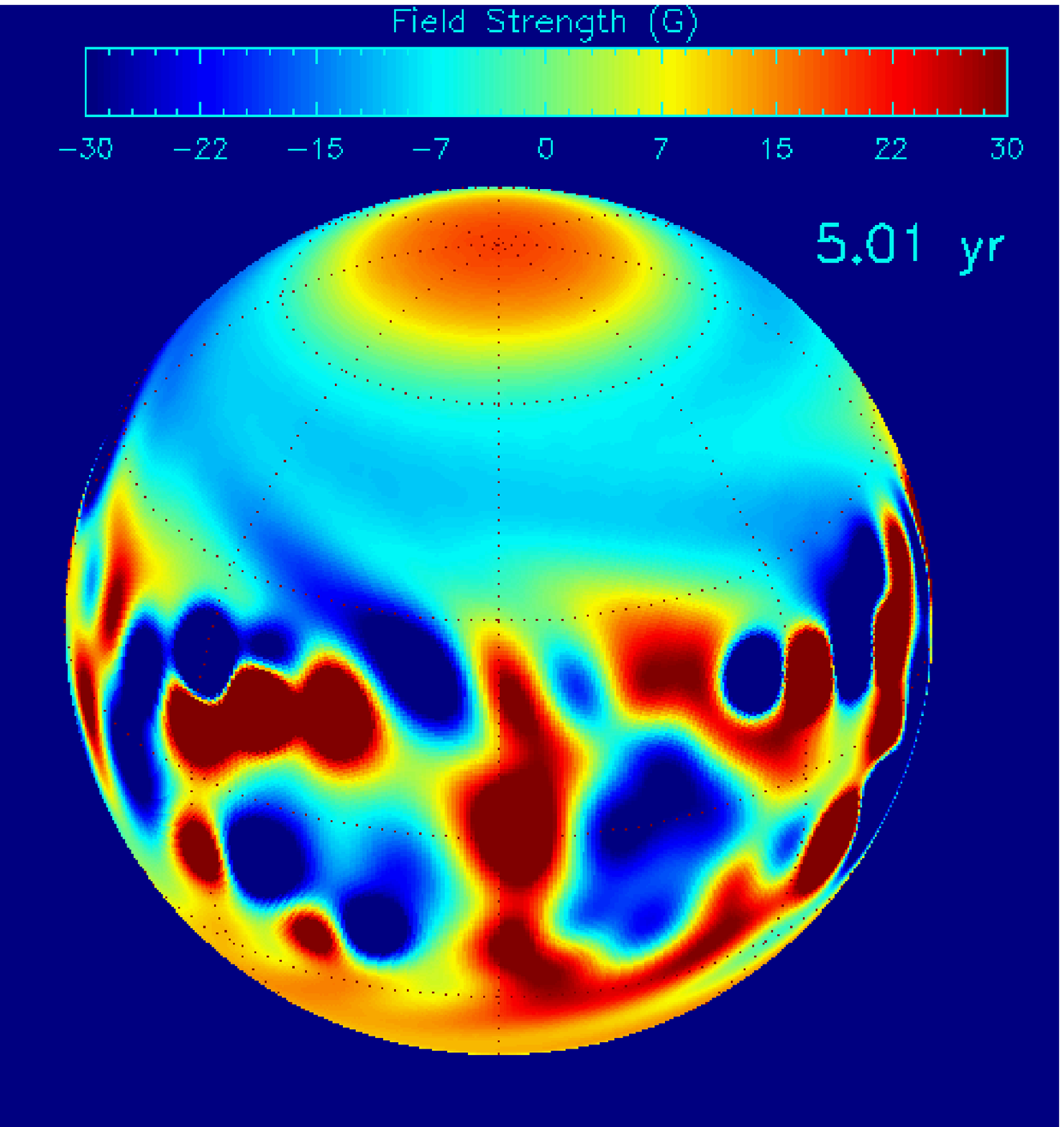}\hskip-1.5mm\\
\vskip2mm
\includegraphics[width=.5\linewidth]{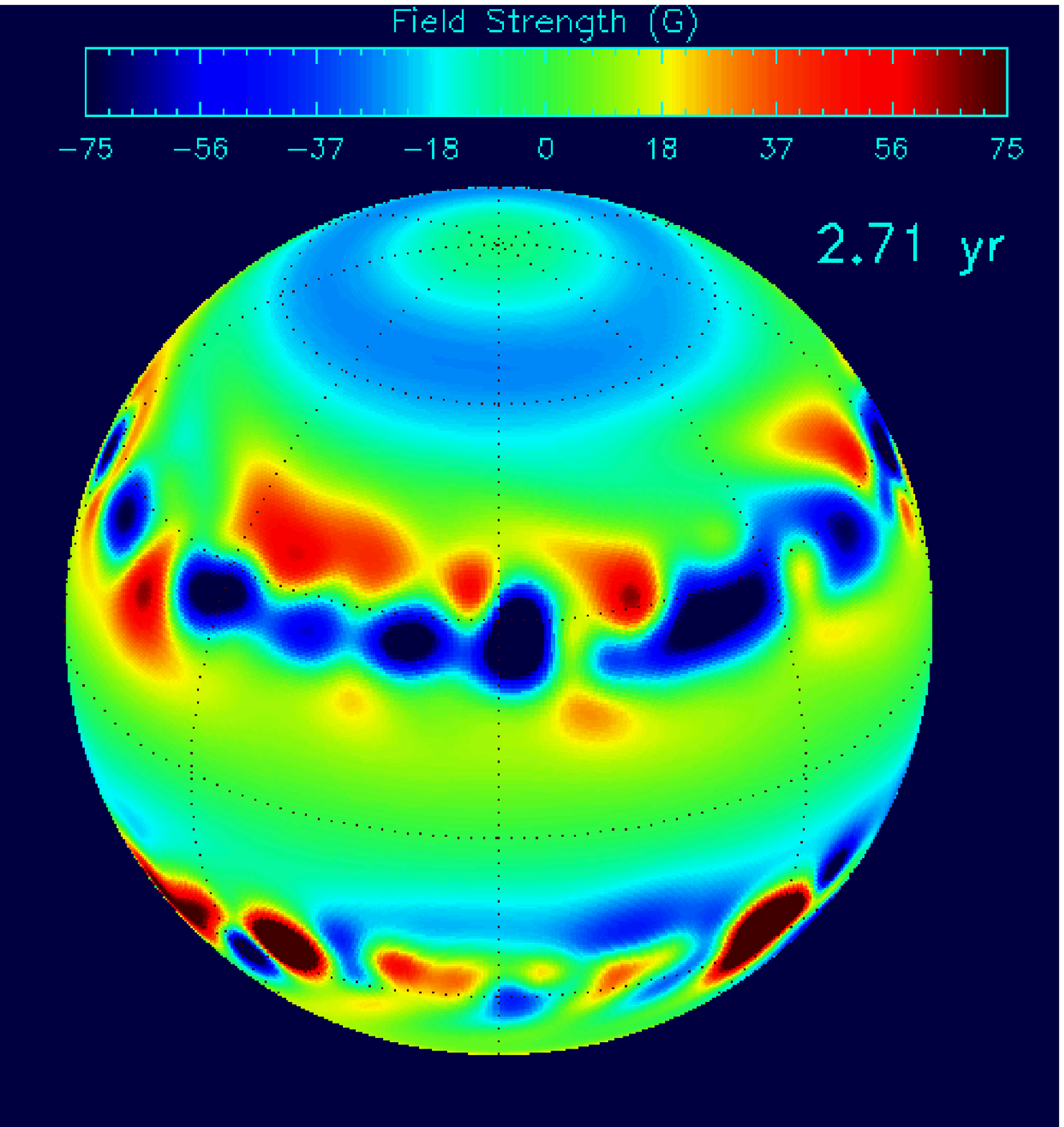}\hskip-1.5mm
\includegraphics[width=.5\linewidth]{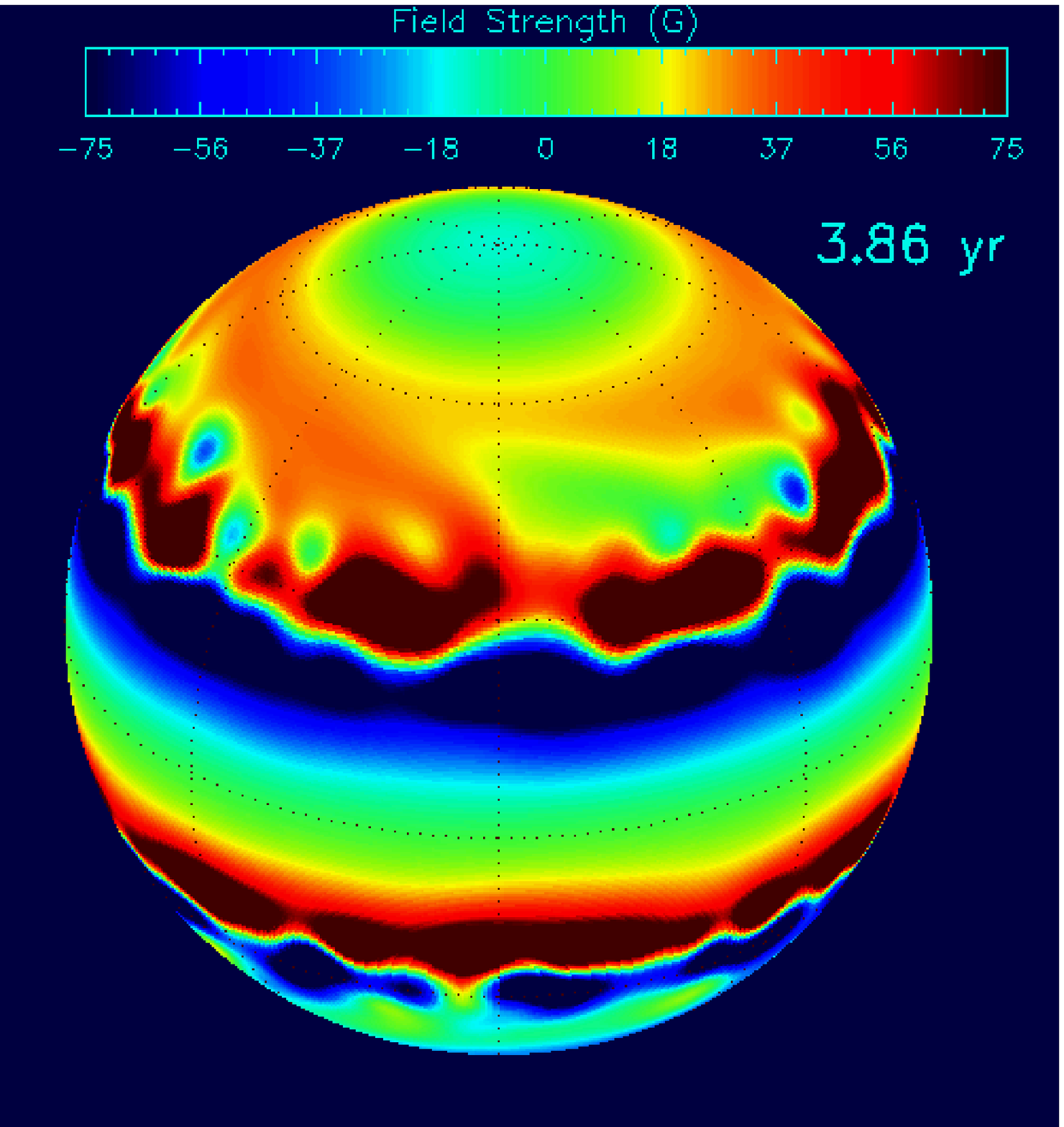}\hskip-1.5mm
\caption{The surface distributions of magnetic flux for the Sun-like star 
with $P_{\rm rot}=27$~d (top panels), $P_{\rm rot}=9$~d (middle panels), 
and $P_{\rm rot}=2$~d (bottom panels), near the activity minimum (left panels) 
and the maximum phases (right panels). 
The colour scale for the magnetic field strength 
saturates at $\pm 5,30$, and $75$~G for top, middle, and bottom panels, respectively. 
The corresponding time-latitude diagrams for the azimuthally 
averaged magnetic field strength are shown in Figs.~\ref{fig:sunall}c, 
\ref{fig:9dall}c, and \ref{fig:2dall}c.}
\label{fig:bsurf}
\end{figure}

We provide three animated GIF files showing the simulated evolution 
of surface magnetic flux for three rotation rates of the Sun-like star, 
available on-line. 

The animation {\tt Bsurf\_27d.gif} shows the evolution of magnetic field 
at the surface for the Sun-like model (Sect.~\ref{ssec:size}). 
Still frames from the animation are shown in Fig.~\ref{fig:bsurf} 
(top panels). The simulation begins at a cycle phase close
to the polar field maximum with negative polarity at the north pole. The
overlapping of two activity cycles is visible, i.e., high-latitude BMRs
of the next cycle emerge, while BMRs of the ending cycle continue to
emerge at low latitudes. In the course of the cycle, the polar field is
weakened by cancellation with opposite-polarity (negative) flux
transported from the active lower latitudes and eventually reverses
around activity maximum. Afterwards, BMRs of the new cycle gradually build 
up positive flux at the north pole. 

The animation {\tt Bsurf\_9d.gif} shows the variation in magnetic field 
at the surface of the Sun-like star with $P_{\rm rot}=9$~d 
(Sect.~\ref{sec:10d}). Still frames from the animation are shown in 
Fig.~\ref{fig:bsurf} (middle panels). Although the flux
emerges within a rather restricted latitude range, it is spread over the
entire stellar surface by diffusion and the large-scale flows. There are
unipolar magnetic regions near the poles and mixed-polarity regions at
low to middle latitudes.

The animation {\tt Bsurf\_2d.gif} shows the evolution of magnetic field 
at the surface of the Sun-like star with $P_{\rm rot}=2$~d 
(Sect.~\ref{sec:2d}). Still frames from the animation are shown in 
Fig.~\ref{fig:bsurf} (bottom panels). The latitude range of flux emergence 
is smaller than in the previous cases, and the tilt angle is significantly 
larger. As a result, two bands of magnetic regions of opposite polarity form, 
predominantly during the activity maximum. 

\end{document}